\documentclass[preprint,3p,12pt]{elsarticle}
\usepackage{mathrsfs}
\usepackage{amsmath}
\usepackage{stmaryrd}
\usepackage{bbding}
\usepackage{dcolumn}
\usepackage{graphicx}
\usepackage{amsfonts}
\usepackage{amssymb}
\usepackage{psfrag}
\usepackage{wrapfig}
\usepackage{subfigure}
\usepackage{makeidx}
\usepackage{bm}
\usepackage{epsf}
\usepackage{epsfig}
\usepackage{setspace}
\usepackage{graphicx}
\usepackage{epstopdf}
\usepackage{psfrag}
\usepackage{subfigure}
\usepackage{color}
\usepackage{comment}
\begin{document}

\title{Comparison of the performance of high-order schemes based on the gas-kinetic and HLLC fluxes}

\author[HKUST1]{Xiaojian Yang}
\ead{xyangbm@connect.ust.hk}

\author[HKUST2]{Xing Ji}
\ead{xjiad@connect.ust.hk}

\author[HKUST1]{Wei Shyy}
\ead{weishyy@ust.hk}

\author[HKUST1,HKUST2,HKUST3]{Kun Xu\corref{cor1}}
\ead{makxu@ust.hk}

\address[HKUST1]{Department of Mechanical and Aerospace Engineering, Hong Kong University of Science and Technology, Clear Water Bay, Kowloon, Hong Kong}
\address[HKUST2]{Department of Mathematics, Hong Kong University of Science and Technology, Clear Water Bay, Kowloon, Hong Kong}
\address[HKUST3]{Shenzhen Research Institute, Hong Kong University of Science and Technology, Shenzhen, China}
\cortext[cor1]{Corresponding author}

\begin{abstract}
In this paper, a comparison of the performance of two high-order finite volume methods based on the gas-kinetic scheme (GKS) and HLLC fluxes is carried out in structured rectangular mesh. For both schemes, the fifth-order WENO-AO reconstruction is adopted to achieve a high-order spatial accuracy. In terms of temporal discretization, a two-stage fourth-order (S2O4) time marching strategy is adopted for WENO5-AO-GKS scheme,
and the fourth-order Runge-Kutta (RK4) method is employed for WENO5-AO-HLLC scheme. For the viscous flow computation, the GKS includes both  inviscid and viscous fluxes in the evolution of a single cell interface gas distribution function. While for the WENO5-AO-HLLC scheme, the inviscid flux is provided by HLLC Riemann solver, and the viscous flux is discretized by a sixth-order central difference method.
Based on the tests of forward Mach step and viscous shock tube, both schemes show outstanding shock capturing property.
From the Titarev-Toro and double shear layer tests, WENO5-AO-GKS scheme seems to have a better resolution than WENO5-AO-HLLC scheme.
Both schemes show excellent robustness in extreme cases, such as the Le Blanc problem. From the cases of the Noh problem and the compressible isotropic turbulence, WENO5-AO-GKS scheme shows favorite robustness. In the compressible isotropic turbulence and three-dimensional Taylor-Green vortex problems, WENO-AO-GKS can use a CFL number up to $0.5$, instead of $0.3$ for WENO5-AO-HLLC. In terms of computational efficiency, WENO5-AO-HLLC scheme is about 27\% more expensive than WENO5-AO-GKS scheme in the two-dimensional viscous flow problems, but is about 15\% faster in the three-dimensional case, because WENO5-AO-GKS scheme needs multidimensional spatial reconstruction for flow variables in both one normal and two tangential directions in the 3D case. Due to the multi-dimensionality, WENO5-AO-GKS scheme performs better than WENO5-AO-HLLC scheme in the laminar boundary layer and the double shear layer test.
\end{abstract}


\maketitle

\section{Introduction}	
The development of high-order schemes has been the main research direction in the current computational fluid dynamics.
The targeting scheme should be accurate, robust, and efficient.
The finite volume scheme is mainly composed of spatial reconstruction, flux evaluation, and temporal discretization.
The successful high-order reconstructions include the essentially non-oscillatory (ENO)  and weighted essentially non-oscillatory (WENO) scheme \cite{eno-Harten1986, weno,weno-liu1994}. There exists many modified versions of WENO, such as WENO-JS \cite{weno}, WENO-Z \cite{wenoz}, central WENO (CWENO) \cite{centralweno-levy1999}, WENO with adaptive order (WENO-AO) \cite{wenoao-balsara2016efficient}, multi-resolution WENO \cite{zhu2018multiresolution}, etc.

Besides the importance of initial reconstruction, the flux evaluation and temporal updating method also play important roles in the determination of the quality of the schemes.
In the past decades, the gas-kinetic scheme (GKS) is mainly focusing on the time accurate flux function for capturing the Euler and Navier-Stokes solutions. The GKS is based on the kinetic Bhatnagar-Gross-Krook (BGK) model and the Chapman-Enskog expansion is used for the flux evaluation \cite{BGK,CE-expansion}.
The scheme has been systematically developed for the flow computation from low-speed to hypersonic one \cite{GKS-lecture, GKS-2001}.
In GKS, a time-dependent gas distribution function at the cell interface is obtained and covers a physical process from the kinetic free particle transport to the hydrodynamic NS wave propagation.
In the smooth region, GKS can accurately recover the Euler or Navier-Stokes solution. In the discontinuity region, the particle free transport
mechanism introduces the numerical dissipation within a shock layer and stabilize the numerical shock structure.
Different from the traditional CFD methods based on the macroscopic government equations directly, GKS has multiscale property. Depending on the ratio of time step $\Delta t$ over the particle collision time $\tau$, the flux function in GKS makes a smooth transition from the upwind flux vector splitting (kinetic scale) to the central difference (hydrodynamic scale).
GKS has been adopted in multicomponent flow \cite{GKS-multicomponent-xu1997, GKS-multicomponent-pan2017}, acoustic computation \cite{GKS-acoustic-zhao2019}, turbulence simulation \cite{GKS-turbulence-liao2009, GKS-turbulence-implicitHGKS-Cao2019, GKS-turbulence-TanShuang2018}, and hypersonic flow \cite{GKS-hypersonic-LiQibing2005}, etc.
Furthermore, a unified GKS (UGKS) has been developed for all flow regimes from rarefied to continuum one \cite{UGKS}.
At the same time, in order to develop high-order GKS, many techniques in CFD have been used in the kinetic schemes.
The WENO reconstruction has been adopted to improve spatial accuracy \cite{3rdGKS-Luo}.
Also, the high-order compact GKS on both structured and unstructured meshes have been developed \cite{CompactGKS-ji2018-structured, CompactGKS-ji2020-unstructured, CompactGKS-zhao2019-8th-order}.
Since the flux function in GKS is time-dependent, which provides not only the numerical flux but also its time derivative. 
Therefore, multi-stage multi-derivative (MSMD) methods can be employed for time marching in GKS \cite{MSMD-ji2018}.
Particularly, a two-stage fourth-order (S2O4) temporal discretization for GKS has been developed with favorable numerical performance \cite{S2O4-Pan2016, S2O4-threedimensional-Pan2018}.

In the CFD community, mostly the exact or approximate Riemann problems are used in the flux construction \cite{Godunov1959}.
One of the outstanding approximate Riemann solvers is the HLL flux \cite{HLL1983}.
In HLL, a configuration including two waves and three constant states is assumed.
In order to improve the capacity of capturing contact surfaces in HLL solver,
Toro presented a modified version of HLL-type Riemann solver, which was called Harten-Lax-van Leer contact (HLLC), to resolve the contact discontinuity in wave structure and show better resolution of intermediate waves \cite{HLLC1994-Toro}.
In HLLC solver, the priori estimate of the fastest and slowest wave emerging from the initial discontinuity is needed, and several methods have been proposed \cite{Toro2013book}.
Since the HLLC flux is time-independent, the Runge-Kutta (RK) method is usually employed for updating the solution in time.
HLLC Riemann solver has been successfully used in the simulation of two-phase flow \cite{HLLC-twophaseflow-TokarevaToro-2010}, combustion \cite{HLLC-turbulencecombustion-2013}, turbulence \cite{HLLC-turbulence-Batten1997}, etc.
More details and extensions of the HLLC Riemann solver be found in the review paper under the finite volume and discontinuous Galerkin frameworks \cite{HLLC-2019review-Toro}.

There are differences between GKS and Riemann solver based schemes.
In GKS, the inviscid and viscous terms are coupled together in the flux evaluation from a time-dependent gas distribution function,
where the spatial derivatives in the normal and tangential directions are included in the time evolution of the gas distribution function.
The current study is to make a comparison of the performance in inviscid and viscous flow simulations between GKS and HLLC Riemann solver
in terms of accuracy, robustness, efficiency, and stability.
The same fifth-order WENO-AO reconstruction is employed to minimize the differences in spatial discretization for these two schemes.
In WENO-AO reconstruction, both the point-wise quantities and the corresponding spatial derivatives are provided as the initial state \cite{wenoao-gks-ji2019-performance-enhancement}.
Besides, S2O4 temporal discretization is used for GKS, and fourth-order Runge-Kutta (RK4) is adopted for HLLC solver, while both time marching schemes achieve the same temporal accuracy.
For convenience, the above two schemes are named as WENO5-AO-GKS and WENO5-AO-HLLC schemes.

This paper is organized as follows.
In Section 2, the WENO-AO reconstruction, GKS, and HLLC Riemann solver are introduced.
Section 3 presents the simulation results of many test cases by WENO5-AO-GKS and WENO5-AO-HLLC schemes.
Section 4 provides the computational efficiency of these two schemes.
The last section is the conclusion.

\section{WENO-AO-GKS and WENO-AO-HLLC}	

\subsection{WENO-AO reconstruction}	
The WENO-AO reconstruction was proposed by Balsara et al. \cite{wenoao-balsara2016efficient}.
To meet the requirement for a fourth-order scheme in both space and time, the fifth-order WENO5-AO reconstruction is selected.
 Assume that $\overline{Q}$ is the cell-averaged variable, and $Q$ is the reconstructed variable. Both $\overline{Q}$ and $Q$ can be conservative or characteristic variables.
 To achieve fifth-order spatial accuracy for $Q$, three sub-stencils $S_k$, $k=0,1,2$ are used to reconstruct the left $Q_{i+1/2}^l$ and right $Q_{i-1/2}^r$ interface values at $x_{i-1/2}$ and $x_{i+1/2}$. These three sub-stencils $S_k$ are,
\begin{align*}
S_0=\{I_{i-2},I_{i-1}, I_i\}, ~~S_1=\{I_{i-1},I_i, I_{i+1}\}, ~~S_2=\{ I_i,
I_{i+1},I_{i+2}\}.
\end{align*}
For each sub-stencil $S_k$, a unique quadratic polynomial
$p^{r3}_{k}(x)$ is constructed by the requirements,
\begin{align}
\frac{1}{\Delta x}\int_{I_{i-j-k-1}}p^{r3}_{k}(x) \text{d}x=\overline{Q}_{i-j-k-1},~j=-1,0,1,
\end{align}
and each $p^{r3}_{k}(x)$ achieves a third-order spatial accuracy in smooth flow region.

On a large stencil $\mathbb{S}_3 = \{S_0, S_1, S_2\}$,
a unique fourth-order polynomial $p_3^{r5}(x)$ is obtained by
\begin{align*}
\frac{1}{\Delta x}\int_{I_{i+j}}p_3^{r5}(x)\text{d}x=\overline{Q}_{i+j}, ~j=-2,-1, 0,1,2.
\end{align*}

After determining the above reconstructions based on different stencils, $p_3^{r5}(x)$ is defined again as,
\begin{align}\label{rewrite-r5}
p_{3}^{r5}(x)&=\gamma_3[\frac{1}{\gamma_3}p_3^{r5}(x)-\sum_0^2 \frac{\gamma_k}{\gamma_3}p_k^{r3}(x)]+\sum_0^2 {\gamma_k}p_k^{r 3}(x),
\end{align}
where $\gamma_{k}$, $k=0,1,2,3$ are linear weights. According to Balsara et al. \cite{wenoao-balsara2016efficient}, the coefficients are given by
\begin{align*}
\gamma_3=\gamma_{Hi},~~ \gamma_0=\gamma_2=(1-\gamma_{Hi})(1-\gamma_{Lo})/2, ~~\gamma_1=(1-\gamma_{Hi})\gamma_{Lo},
\end{align*}
where $\gamma_{Hi} \in [0.85,0.95]$ and $\gamma_{Lo} \in [0.85,0.95]$.
Obviously, the above formulas satisfy $\sum_0^3 \gamma_{k} = 1$ and $\gamma_{k}>0, k=0,1,2,3$.
In the current study, $\gamma_{Hi} =0.85 $ and $\gamma_{Lo} =0.85 $ are used.

To deal with discontinuities, the WENO-Z type \cite{wenoz} non-linear weights are adopted,
\begin{align*}
\omega_k=\gamma_k[ 1+ \frac{\delta^2}{(\beta_{k}+\epsilon)^2} ],
\end{align*}
where $\delta$ is the global smooth indicator, and it is defined as
\begin{align*}
\delta=\frac{1}{3}(|\beta_3^{r5}-\beta_0^{r3}|+|\beta_3^{r5}-\beta_1^{r3}|+|\beta_3^{r5}-\beta_2^{r3}|) = O(\Delta h^4).
\end{align*}
More specifically, $\beta_{k}=\beta_k^{r3}$, $k=0,1,2$, are the smooth indicator of sub-stencil $S_k$, and $\beta_3=\beta_3^{r5}$ is the smooth indicator of the large stencil $\mathbb{S}_3$. The explicit formulas of $\beta_{k}$ can refer to \cite{wenoao-balsara2016efficient}. Besides, $\epsilon$ is a positive small number to avoid zero for denominator with $\epsilon = 10^{-8}$.
Then, the normalized weights $\overline{\omega}_k$ can be defined as follows,
\begin{align*}
\overline{\omega}_k=\frac{\omega_k}{\sum_0^3 \omega_q}.
\end{align*}
The final form of the reconstructed polynomial can be written as,
\begin{align}\label{weno-ao-re-polynomial}
P^{AO(5,3)}(x)=\overline{\omega}_3[\frac{1}{\gamma_3}p_3^{r5}(x)-\sum_0^2 \frac{\gamma_k}{\gamma_3}p_k^{r3}(x)]+\sum_0^2 {\overline{\omega}_k}p_k^{r 3}(x).
\end{align}
The reconstructed left interface value $Q_{i+1/2}^l$ and the corresponding derivative become,
\begin{align*}
Q^l_{i+1/2}=P^{AO(5,3)}(x_{i+1/2}), ~~(Q^l_{x})_{i+1/2}=P_x^{AO(5,3)}(x_{i+1/2}).
\end{align*}
Similarly, the right interface value $Q^r_{i-1/2}$ and its derivative can also be determined by,
\begin{align*}
Q^r_{i-1/2}=P^{AO(5,3)}(x_{i-1/2}), ~~ (Q^r_{x})_{i-1/2}=P_x^{AO(5,3)}(x_{i-1/2}).
\end{align*}

The reconstructed value and its normal derivative at the Gaussian quadrature points are obtained from the above procedure.
Since GKS needs not only normal derivatives $\left(Q_x\right)$, but also tangential derivative $\left(Q_y, Q_z\right)$,
the multi-dimensional reconstruction is performed. The details are given in \cite{wenoao-gks-ji2019-performance-enhancement}.

\subsection{WENO5-AO-GKS scheme}
\subsubsection{BGK equation and gas-kinetic scheme}	
Here the GKS in 2D case is presented and the scheme in 3D can be obtained similarly.
The two-dimensional BGK equation is 
written as \cite{BGK},
\begin{equation}\label{bgk}
f_t+\textbf{u}\cdot\nabla f=\frac{g-f}{\tau},
\end{equation}
where $\textbf{u}$ is the particle velocity, $f$ is the gas distribution function, $g$ is the corresponding equilibrium state, and $\tau$ is the collision time. The collision term satisfies the compatibility condition
\begin{equation}\label{equ_compatibility}
\int \frac{g-f}{\tau} \bm{\psi} \text{d}\Xi=0,
\end{equation}
where $\bm{\psi}=(1,u,v,\displaystyle \frac{1}{2}(u^2+v^2+\xi^2))^T$, the internal variables $\xi^2=\xi_1^2+...+\xi_K^2$, $\text{d}\Xi=\text{d}u\text{d}v\text{d}\xi_1...\text{d}\xi_{K}$, $K$
is the internal degree of freedom, i.e.
$K=(4-2\gamma)/(\gamma-1)$ for two-dimensional flows, and $\gamma$
is the specific heat ratio.

In the continuum regime, the gas distribution function can be expanded as
\begin{align*}
f=g-\tau D_{\textbf{u}}g+\tau D_{\textbf{u}}(\tau
D_{\textbf{u}})g-\tau D_{\textbf{u}}[\tau D_{\textbf{u}}(\tau
D_{\textbf{u}})g]+...,
\end{align*}
where $D_{\textbf{u}}={\partial}/{\partial t}+\textbf{u}\cdot
\nabla$. The corresponding macroscopic equations can be derived by truncating on different order of $\tau$. For example, when the zeroth-order truncation is taken, i.e. $f=g$, the Euler equations can be derived. When the first-order truncation is used,
\begin{align}\label{ns}
f=g-\tau (ug_x+vg_y+g_t),
\end{align}
the Navier-Stokes equations can be derived with $\tau = \mu /p $. The difficulties for the development of a reliable gas-kinetic scheme is the possible
discontinuity of flow variables at the cell interface, where the above Chapman-Enskog expansion cannot be used directly for the flux evaluation,
 and the time evolution solution of the gas distribution function at the cell interface has to be constructed properly from a piecewise
 discontinuous initial condition.  

Based on the conservation laws in a discretized space of control volume $S_{ij}=\left[x_i-\Delta x/2,x_i+\Delta x/2\right] \times \left[y_j-\Delta y/2,y_j+\Delta y/2\right]$, the semi-discrete form of finite volume scheme can be obtained as
\begin{align}\label{equ_semi_discrete}
\frac{\text{d}\textbf{W}_{ij}}{\text{d}t}=-\frac{1}{\Delta x}
(\textbf{F}_{i+1/2,j}(t)-\textbf{F}_{i-1/2,j}(t))-\frac{1}{\Delta y}
(\textbf{G}_{i,j+1/2}(t)-\textbf{G}_{i,j-1/2}(t)),
\end{align}
where $\textbf{W}_{ij}=\left[\rho,\rho U, \rho V, \rho E\right]^T$ are the cell-averaged conservative variables. $\textbf{F}_{i\pm 1/2,j}(t)$ and $\textbf{G}_{i,j \pm 1/2}(t)$ are the time-dependent numerical fluxes across the cell interfaces in $x$ and $y$ directions respectively.
  The fluxes can be obtained by a time-dependent gas distribution function $f$ at the corresponding cell interface. To achieve the accuracy in space, the Gaussian quadrature is used. Taking the numerical fluxes in $x$ directions $\textbf{F}_{i+1/2,j}(t)$, for example,
\begin{align}\label{equ_gauss}
\textbf{F}_{i+1/2,j}(t)=\frac{1}{\Delta
	y}\int_{y_{j-1/2}}^{y_{j+1/2}}\textbf{F}_{i+1/2}(y,t)\text{d}y=\sum_{\ell=1}^2\omega_\ell \textbf{F}_{i+1/2,j_\ell}(t),
\end{align}
two Gaussian quadrature points $\displaystyle y_{j_\ell}=y_j+\frac{(-1)^{\ell-1}}{2\sqrt{3}}\Delta y$, $\ell= 1, 2$, and the corresponding weights $\omega_1=\omega_2=1/2$ are employed in this paper, which yields fourth-order accuracy in space. $\textbf{F}_{i+1/2,j_\ell}(t)$, $\ell= 1, 2$, are numerical fluxes at the Gaussian quadrature points,
\begin{align}
\textbf{F}_{i+1/2,j_\ell}(t)=\int\bm{\psi} u f(x_{i+1/2},y_\ell,t,u,v,\xi)\text{d}\Xi,
\end{align}
where $f(x_{i+1/2},y_\ell,t,u,v,\xi)$, $\ell= 1, 2$, are the gas distribution function at the Gaussian points. To obtain the numerical fluxes, the integral solution of BGK equation Eq.\eqref{bgk} at point $(x_{i+1/2},y_\ell)$ and time $t$ is used,
\begin{equation}\label{equ_integral1}
f(x_{i+1/2},y_\ell,t,u,v,\xi)=\frac{1}{\tau}\int_0^t g(x',y',t',u,v,\xi)e^{-(t-t')/\tau}\text{d}t'\\
+e^{-t/\tau}f_0(-ut,-vt,u,v,\xi),
\end{equation}
where $(x_{i+1/2}, y_\ell)=(0,0)$ for the simplification of the notation, $x=x'+u(t-t')$ and $y=y'+v(t-t')$ are the trajectory of particles. $f_0$ is the initial gas distribution function at time $t=0$, and $g$ is the corresponding equilibrium state.

In the integral solution Eq.\eqref{equ_integral1},
the initial gas distribution function can be constructed as
\begin{equation}\label{equ_f0}
f_0=f_0^l(x,y,u,v)H(x)+f_0^r(x,y,u,v)(1-H(x)),
\end{equation}
where $H(x)$ is the Heaviside function, $f_0^l$ and $f_0^r$ are the
initial gas distribution functions on the left and right side of one cell
interface, which can be determined by the corresponding macroscopic variables.
The initial gas distribution function $f_0^k$, $k=l,r$, is constructed as
\begin{equation*}
f_0^k=g^k\left(1+a^kx+b^ky-\tau(a^ku+b^kv+A^k)\right),
\end{equation*}
where $g^l$ and $g^r$ are the Maxwellian distribution functions on the left and right hand sides of a cell interface, and they can be determined by the corresponding conservative variables $\textbf{W}^l$ and $\textbf{W}^r$. The coefficients $a^l$, $a^r$, $b^l$, $b^r$ are related to the spatial derivatives in normal and tangential directions, which can be obtained from the corresponding derivatives of the initial macroscopic variables,
\begin{equation*}
\left\langle a^l\right\rangle=\partial \textbf{W}^l/\partial x,
\left\langle a^r\right\rangle=\partial \textbf{W}^r/\partial x,
\left\langle b^l\right\rangle=\partial \textbf{W}^l/\partial y,
\left\langle b^r\right\rangle=\partial \textbf{W}^r/\partial y,
\end{equation*}
where $\left\langle...\right\rangle$ means the moments of the Maxwellian distribution function,
\begin{align*}
\left\langle...\right\rangle=\int \bm{\psi}\left(...\right)g\text{d}\Xi.
\end{align*}
The non-equilibrium parts on the Chapman-Enskog expansion have no net contribution to the conservative variables,
\begin{equation*}
\left\langle a^lu+b^lv+A^l\right\rangle = 0,~
\left\langle a^ru+b^rv+A^r\right\rangle = 0,
\end{equation*}
and therefore the coefficients $A^l$ and $A^r$, related to time derivatives, can be obtained.
After the determination of $f_0$, the equilibrium state $g$ around the cell interface is modeled as,
\begin{equation}\label{equ_g}
g=g_0\left(1+\overline{a}x+\overline{b}y+\bar{A}t\right),
\end{equation}
where $g_0$ is the local equilibrium at point $(x_{i+1/2},y_\ell)$ and can be determined by the compatibility condition,
\begin{align}
\int\bm{\psi} g_{0}\text{d}\Xi=\textbf{W}_0
&=\int_{u>0}\bm{\psi} g^{l}\text{d}\Xi+\int_{u<0}\bm{\psi} g^{r}\text{d}\Xi, \nonumber \\
\int\bm{\psi} \overline{a} g_{0}\text{d}\Xi=\partial \textbf{W}_0/\partial x
&=\int_{u>0}\bm{\psi} a^l g^{l}\text{d}\Xi+\int_{u<0}\bm{\psi} a^r g^{r}\text{d}\Xi,\\
\int\bm{\psi} \overline{b} g_{0}\text{d}\Xi=\partial \textbf{W}_0/\partial y
&=\int_{u>0}\bm{\psi} b^l g^{l}\text{d}\Xi+\int_{u<0}\bm{\psi} b^r g^{r}\text{d}\Xi, \nonumber
\end{align}
and
\begin{equation*}
\left\langle \overline{a}u+\overline{b}v+\bar{A}\right\rangle = 0.
\end{equation*}
After constructing the initial gas distribution function $f_0$ and the equilibrium state $g$, and substituting Eq.\eqref{equ_f0} and Eq.\eqref{equ_g} into Eq.\eqref{equ_integral1}, the time-dependent distribution function $f(x_{i+1/2}, y_\ell, t,u,v,\xi)$ at a cell interface can be expressed as,
\begin{align}\label{equ_finalf}
f(x_{i+1/2,j_\ell},t,u,v,\xi)=&(1-e^{-t/\tau})g_0+[(t+\tau)e^{-t/\tau}-\tau](\overline{a}u+\overline{b}v)g_0\nonumber\\
+&(t-\tau+\tau e^{-t/\tau}){\bar{A}} g_0\nonumber\\
+&e^{-t/\tau}g^r[1-(\tau+t)(a^ru+b^rv)-\tau A^r]H(u)\nonumber\\
+&e^{-t/\tau}g^l[1-(\tau+t)(a^lu+b^lv)-\tau A^l](1-H(u)).
\end{align}

The collision time $\tau$  in Eq.\eqref{equ_finalf} is defined by
\begin{align*}
\tau=\frac{\mu}{p}+c_2 \displaystyle|\frac{p_l-p_r}{p_l+p_r}|\Delta t,
\end{align*}
for viscous flow computation, where $p_l$ and $p_r$ are the pressures on the left and right
sides of the cell interface, and $p$ is the pressure at the interface from the equilibrium state. 
Here $\Delta t$ is the time step.
For inviscid flow, the $\tau$ is given by
\begin{align*}
\tau=c_1 \Delta t+c_2\displaystyle|\frac{p_l-p_r}{p_l+p_r}|\Delta t,
\end{align*}
where $c_1=0.01$, $c_2=1\sim5$.

\subsubsection{Two-stage fourth-order temporal discretization}
The two-stage fourth-order temporal discretization, originally developed for the generalized Riemann problem (GRP) solver \cite{S2O4-li2016},
   has been applied to GKS \cite{S2O4-Pan2016}. A fourth-order time-accurate GKS can be constructed using the second-order flux function Eq.\eqref{equ_finalf}.
For the time-dependent equations,
\begin{align}\label{pde}
\frac{\partial \textbf{W}}{\partial t}=\mathcal {L}(\textbf{W}),
\end{align}
with the initial condition at $t_n$,
\begin{align}\label{pde2}
\textbf{W}(t=t_n)=\textbf{W}^n,
\end{align}
where $\mathcal {L}$ is the spatial operator of flux obtained in Eq.\eqref{equ_semi_discrete},
a fourth-order temporal accurate solution for $\textbf{W}(t)$ at $t=t_n +\Delta t$ can be updated by,
\begin{align}\label{s2o4-1}
\textbf{W}^*=\textbf{W}^n+\frac{1}{2}\Delta t\mathcal{L}(\textbf{W}^n)
+\frac{1}{8}\Delta t^2\frac{\partial}{\partial t}\mathcal{L}(\textbf{W}^n).
\end{align}
\begin{align}\label{s2o4-2}
\textbf{W}^{n+1}=\textbf{W}^n+\Delta t\mathcal{L}(\textbf{W}^n)
+\frac{1}{6}\Delta t^2\big(\frac{\partial}{\partial t}\mathcal{L}(\textbf{W}^n)
+2\frac{\partial}{\partial t}\mathcal{L}(\textbf{W}^*)\big).
\end{align}
The detailed proof is given in \cite{S2O4-li2016}.

The numerical fluxes and their time derivatives in the above equations,
such as $\mathcal{L}(W_i^n)$ and $\frac{\partial}{\partial t}\mathcal{L}(W_i^n)$, are determined by
\begin{equation}\label{flux-operator-2d-tn}
\begin{split}
\mathcal{L}(\textbf{W}_{i,j}^n)=
&-\frac{1}{\Delta x}
[(\textbf{F})_{i+1/2,j}(\textbf{W}^n,t_n)-(\textbf{F})_{i-1/2,j}(\textbf{W}^n,t_n)] \\
&-\frac{1}{\Delta y}
[(\textbf{G})_{i,j+1/2}(\textbf{W}^n,t_n)-(\textbf{G})_{i,j-1/2}(\textbf{W}^n,t_n)], \\
\mathcal{L}_{t}(\textbf{W}_{i,j}^n)=
&-\frac{1}{\Delta x}
[{\partial_t}(\textbf{F})_{i+1/2,j}(\textbf{W}^n,t_n)-{\partial_t}(\textbf{F})_{i-1/2,j}(\textbf{W}^n,t_n)] \\
&-\frac{1}{\Delta y}
[{\partial_t}(\textbf{G})_{i,j+1/2}(\textbf{W}^n,t_n)-{\partial_t}(\textbf{G})_{i,j-1/2}(\textbf{W}^n,t_n)].
\end{split}
\end{equation}
Similarly, the time derivatives for the intermediate state can be obtained,
\begin{equation}\label{flux-operator-2d-t*}
\begin{split}
\mathcal{L}_{t}(\textbf{W}_{i,j}^*)=
&-\frac{1}{\Delta x}
[{\partial_t}(\textbf{F})_{i+1/2,j}(\textbf{W}^*,t_*)-{\partial_t}(\textbf{F})_{i-1/2,j}(\textbf{W}^*,t_*)] \\
&-\frac{1}{\Delta y}
[{\partial_t}(\textbf{G})_{i,j+1/2}(\textbf{W}^*,t_*)-{\partial_t}(\textbf{G})_{i,j-1/2}(\textbf{W}^*,t_*)].
\end{split}
\end{equation}
In the gas-kinetic scheme, the flux Eq.\eqref{equ_gauss} is a complicated function of time. To obtain the time derivatives of the flux function used in the above two-stage fourth-order framework, the flux function is approximated as a linear function of time within a time interval. The time-dependent flux can be expanded as,
\begin{align}\label{expansion}
\textbf{F}_{i+1/2,j}(\textbf{W}^n,t)=\textbf{F}_{i+1/2,j}^n+\partial_t \textbf{F}_{i+1/2,j}^n \left(t-t_n\right), t\in\left[t_n,t_n+\Delta t\right].
\end{align}
To get coefficients $\textbf{F}_{i+1/2,j}^n$ and $\partial_t\textbf{F}_{i+1/2,j}^n$, the following notation of Eq.\eqref{equ_gauss} is introduced,
\begin{align*}
\mathbb{F}_{i+1/2,j}(\textbf{W}^n,\delta)
=\int_{t_n}^{t_n+\delta}\textbf{F}_{i+1/2,j}(\textbf{W}^n,t)\text{d}t&
=\sum_{\ell=1}^2\omega_\ell \int_{t_n}^{t_n+\delta}\int u \bm{\psi} f(x_{i+1/2,j_\ell},t,u, v,\xi)\text{d}\Xi\text{d}t.
\end{align*}
Take $\delta$ as $\Delta t$ and $\Delta t/2$, we have,
\begin{align*}
\textbf{F}_{i+1/2,j}(\textbf{W}^n,t_n)\Delta t&+\frac{1}{2}\partial_t
\textbf{F}_{i+1/2,j}(\textbf{W}^n,t_n)\Delta t^2 =\mathbb{F}_{i+1/2,j}(\textbf{W}^n,\Delta t) ,\\
\frac{1}{2}\textbf{F}_{i+1/2,j}(\textbf{W}^n,t_n)\Delta t&+\frac{1}{8}\partial_t
\textbf{F}_{i+1/2,j}(\textbf{W}^n,t_n)\Delta t^2 =\mathbb{F}_{i+1/2,j}(\textbf{W}^n,\Delta t/2).
\end{align*}
Solving the above linear system, and we can obtain the expression of coefficients
\begin{align*}
\textbf{F}_{i+1/2,j}(\textbf{W}^n,t_n)&=(4\mathbb{F}_{i+1/2,j}(\textbf{W}^n,\Delta t/2)-\mathbb{F}_{i+1/2,j}(\textbf{W}^n,\Delta t))/\Delta t,\\
\partial_t \textbf{F}_{i+1/2,j}(\textbf{W}^n,t_n)&=4(\mathbb{F}_{i+1/2,j}(\textbf{W}^n,\Delta t)-2\mathbb{F}_{i+1/2,j}(\textbf{W}^n,\Delta t/2))/\Delta
t^2.
\end{align*}
Similarly, the coefficients for the intermediate state $\textbf{F}_{i+1/2,j}(\textbf{W}^*,t_*)$, $\partial_t \textbf{F}_{i+1/2,j}(\textbf{W}^*,t_*)$ can be determined as well.
The fluxes in y-direction can be obtained through the same method.
Then, the intermediate states $\textbf{W}^*_{ij}$ are updated by Eq.\eqref{s2o4-1} and Eq.\eqref{flux-operator-2d-tn}.
The final states $\textbf{W}^{n+1}_{ij}$ in Eq.\eqref{s2o4-2} are determined through Eq.\eqref{flux-operator-2d-tn} and Eq.\eqref{flux-operator-2d-t*}.

\subsection{WENO5-AO-HLLC scheme}	

\subsubsection{HLLC Riemann solver}	
The HLLC Riemann solver \cite{HLLC1994-Toro} is used to obtain the inviscid flux in the current WENO5-AO-HLLC scheme. Consider the following Riemann problem,
\begin{equation*}
\textbf{W}_t + \textbf{F}_x\left(\textbf{W}\right) = 0,
\end{equation*}
with the initial condition,
\begin{equation*}
\textbf{W}(x,0)= \left\{\begin{aligned}
&\textbf{W}_L,  & x < 0,\\
&\textbf{W}_R,  & x > 0,
\end{aligned} \right.
\end{equation*}
where $\textbf{W}_L$ and $\textbf{W}_R$ are the initial interface values.  For the two-dimensional Euler equations, the conservative variables $\textbf{W}$ and the corresponding fluxes $\textbf{F}$ are,
\begin{equation*}
\textbf{W}=\left[\rho,~\rho U,~\rho V,~\rho E\right]^T, ~~~~
\textbf{F}=\left[\rho U,~\rho U^2+p,~\rho U V,~U(\rho E+p)\right]^T.
\end{equation*}
HLLC solver is an approximate Riemann solver, which consists of four constant states. Assume that the speeds of the slowest and fastest wave are $S_L$ and $S_R$, and the speed of the middle shear wave is $S_*$. Then, the HLLC solver can be written as follows,
\begin{equation}
\textbf{W}(x,t)= \left\{\begin{aligned}
&\textbf{W}_L,  & \frac{x}{t} \le S_L,\\
&\textbf{W}_{*L},  & S_L \le \frac{x}{t} \le S_{*}, \\
&\textbf{W}_{*R},  & S_{*} \le \frac{x}{t} \le S_R, \\
&\textbf{W}_R,  & \frac{x}{t} \ge S_R,
\end{aligned} \right.
\end{equation}
and the corresponding numerical flux can be defined as,
\begin{equation}
\textbf{F}_{x+1/2}= \left\{\begin{aligned}
&\textbf{F}_L,  & 0 \le S_L,\\
&\textbf{F}_{*L},  & S_L \le 0 \le S_{*}, \\
&\textbf{F}_{*R},  & S_{*} \le 0 \le S_R, \\
&\textbf{F}_R,  & 0 \ge S_R,
\end{aligned} \right.
\end{equation}
where $\textbf{F}_{*K} = \textbf{F}_K + S_L(\textbf{W}_{*K}-\textbf{W}_K ) $, $K=L,R$. The $\textbf{W}_{*K}$, $K=L,R$, is given by,
\begin{equation}
\textbf{W}_{*K} = \rho_K \left(\frac{S_K-U_K}{S_K-S_*} \right)
\begin{bmatrix}
1 \\ S_* \\ V_K \\ \frac{E_K}{\rho_K} + \left(S_*-U_K\right)\left[S_* + \frac{p_K}{\rho_K \left(S_K - U_K\right)}\right]
\end{bmatrix},
\end{equation}
where $S_*$ is related to the speeds $S_L$ and $S_R$, namely
\begin{equation*}
S_*=\frac{p_R-p_L+\rho_L U_L \left(S_L-U_L\right) - \rho_R U_R \left(S_R-U_R\right)}{\rho_L \left(S_L-U_L\right) - \rho_R \left(S_R-U_R\right)}.
\end{equation*}
There are many methods to estimate wave speeds $S_L$ and $S_R$, and a pressure-based wave speed estimate method proposed by Toro is adopted in the current work \cite{Toro2013book}. Firstly, we need to estimate $p_*$, the pressure of the region $x/t \in \left[S_L,S_R\right]$. Based on the Two-Rarefaction Riemann solver (TRRS), the estimated $p_*$ is
\begin{equation*}
p_* = \left[\frac{a_L+a_R-\frac{\gamma-1}{2}\left(U_R-U_L\right)}{a_L/p_L^z + a_R/p_R^z} \right]^{1/z}
\end{equation*}
where $z=\left(\gamma-1\right)/\left(2\gamma\right) $, and $\gamma$ is the specific heat ratio. Then, the speeds $S_L$ and $S_R$ are coming from
the exact wave-speed relations in the exact Riemann solver,
\begin{equation*}
S_L=U_L-a_L q_L, ~~ S_R=U_R-a_R q_R,
\end{equation*}
where $a_L, a_R$ are the sound speeds of initial left and right state, and $q_K$, $K=L,R$, are
\begin{equation*}
q_K= \left\{\begin{aligned}
&1,  & p_* \le p_K,\\
&\left[ 1+\frac{\gamma+1}{2\gamma} \left( p_*/p_K-1\right)\right]^{1/2},  & p_* > p_K.
\end{aligned} \right.
\end{equation*}

\subsubsection{Viscous flux}	
For viscous flow problems, the viscous fluxes in Navier-Stokes equations are needed. To calculate the viscous fluxes in the current WENO5-AO-HLLC scheme, both the conservative variables $Q_{i+1/2}$ and the corresponding derivatives $\left(Q_x\right)_{i+1/2}$ at the cell interface need to be constructed by the cell averaged conservative variables $\overline{Q}$. In this paper, a sixth-order central difference method is applied for the calculation of viscous fluxes. The conservative variables can be written as follows,
\begin{align*}
Q_{i+1/2}=\frac{1}{60}(\overline{Q}_{i-2}-8\overline{Q}_{i-1}+37\overline{Q}_{i}+37\overline{Q}_{i+1}-8\overline{Q}_{i+2}+\overline{Q}_{i+3}),
\end{align*}
and the corresponding derivatives are,
\begin{align*}
\left(Q_x\right)_{i+1/2}=\frac{1}{180\Delta x}(-2\overline{Q}_{i-2}+25\overline{Q}_{i-1}-245\overline{Q}_{i}+245\overline{Q}_{i+1}-25\overline{Q}_{i+2}+2\overline{Q}_{i+3}).
\end{align*}

For two-dimensional problems, the dimension-by-dimension strategy is adopted \cite{accuracy-FVM,wenoao-gks-ji2019-performance-enhancement}.
The reconstructed value $Q_{i+1/2,j_l}$ at the Gaussian quadrature point $j_l$, the corresponding normal derivative $\left(Q_x\right)_{i+1/2,j_l}$, and tangential derivative $\left(Q_y\right)_{i+1/2,j_l}$ can be obtained by the fourth-order polynomial $p^{r5}\left(y\right)$ based on the above $Q_{i+1/2}$ and $\left(Q_x\right)_{i+1/2}$. Then, all terms in the viscous fluxes can be fully determined.
A similar procedure can be easily extended to three-dimensional problems. To improve the robustness of WENO5-AO-HLLC scheme, the conservative variables at the cell interface $Q_{i+1/2}$ are obtained by simple averaging of the left and right interface values of WENO5-AO reconstruction in some challenging cases.

\subsubsection{Time marching method}
Considering the fourth-order temporal accuracy in WENO5-AO-GKS scheme, the classical fourth-order Runge-Kutta method (RK4) is adopted for time integration in WENO5-AO-HLLC scheme for achieving the 4th-order temporal accuracy.
The RK4 time marching method reads,
\begin{align*}
&\textbf{W}^{1}=\textbf{W}^n+ \frac{1}{2} \Delta t\mathcal{L}(\textbf{W}^n), \\
&\textbf{W}^{2}=\textbf{W}^n + \frac{1}{2} \Delta t\mathcal{L}(\textbf{W}^1), \\
&\textbf{W}^{3}=\textbf{W}^n + \Delta t\mathcal{L}(\textbf{W}^2), \\
&\textbf{W}^{n+1}=\textbf{W}^n +\frac{1}{6} \left( \Delta t\mathcal{L}(\textbf{W}^n) + 2\Delta t\mathcal{L}(\textbf{W}^1) + 2\Delta t\mathcal{L}(\textbf{W}^2) + \Delta t\mathcal{L}(\textbf{W}^3)\right),
\end{align*}
with $\mathcal{L}$ defined in Eq.\eqref{flux-operator-2d-tn}.


\section{Numerical performance}

In the following test cases, for the inviscid flow the time step is determined by,
\begin{equation*}
\Delta t = \text{CFL} \times \frac{\Delta x}{(|\textbf{U}|+C)_{\text{Max}}} ,
\end{equation*}
where $C$ is sound speed. For viscous flow, the time step is given by,
\begin{equation*}
\Delta t = \text{CFL} \times \text{Min}\left[\frac{\Delta x}{(|\textbf{U}|+C)_{\text{Max}}},\frac{\rho\Delta x^2}{4\mu}\right] .
\end{equation*}

\subsection{1-D test case}

\subsubsection{Titarev-Toro problem}
\begin{figure}[htbp]
	\centering
	\subfigure{
		\includegraphics[height=6cm]{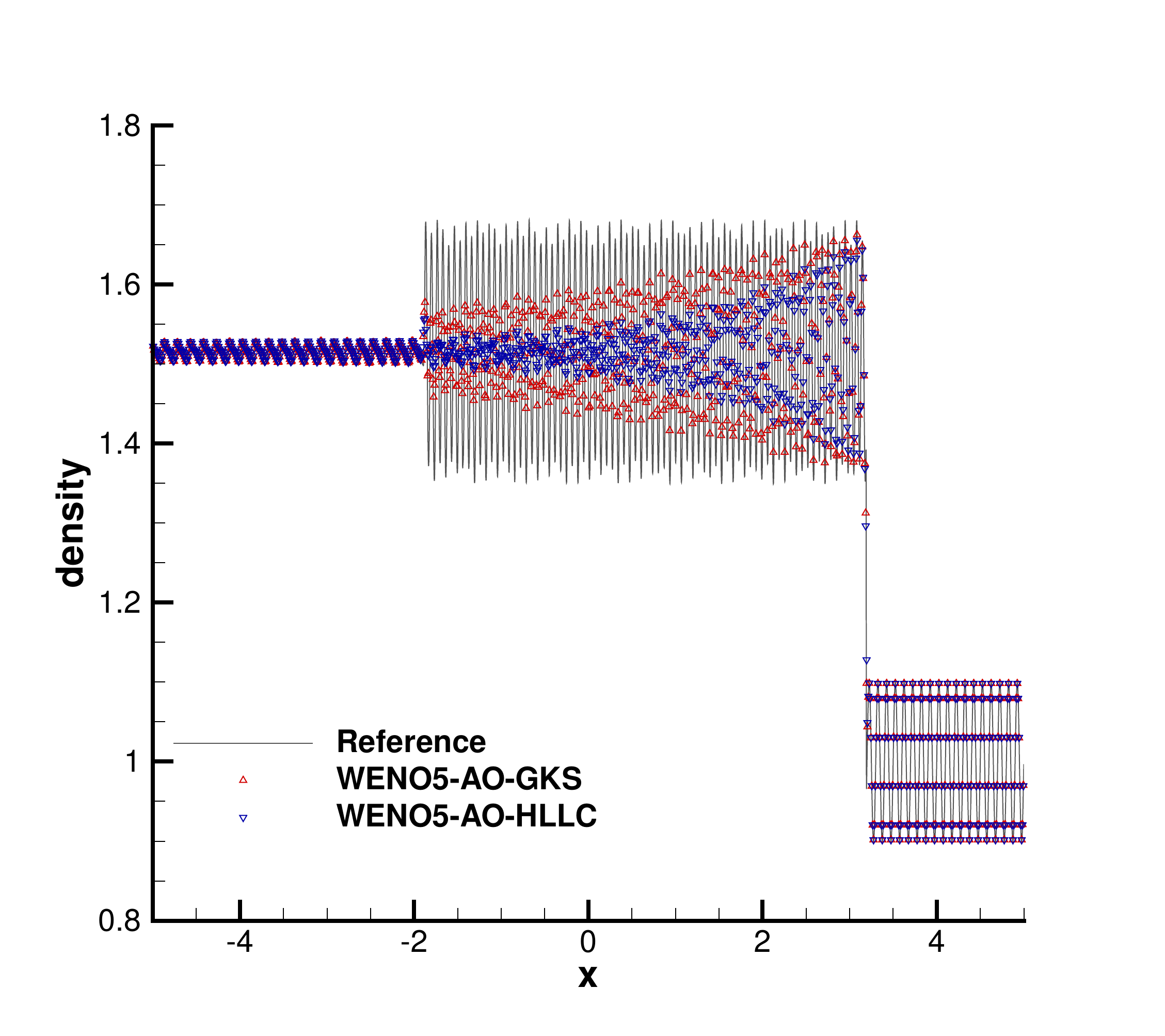}		
	}
	\quad
	\subfigure{
		\includegraphics[height=6cm]{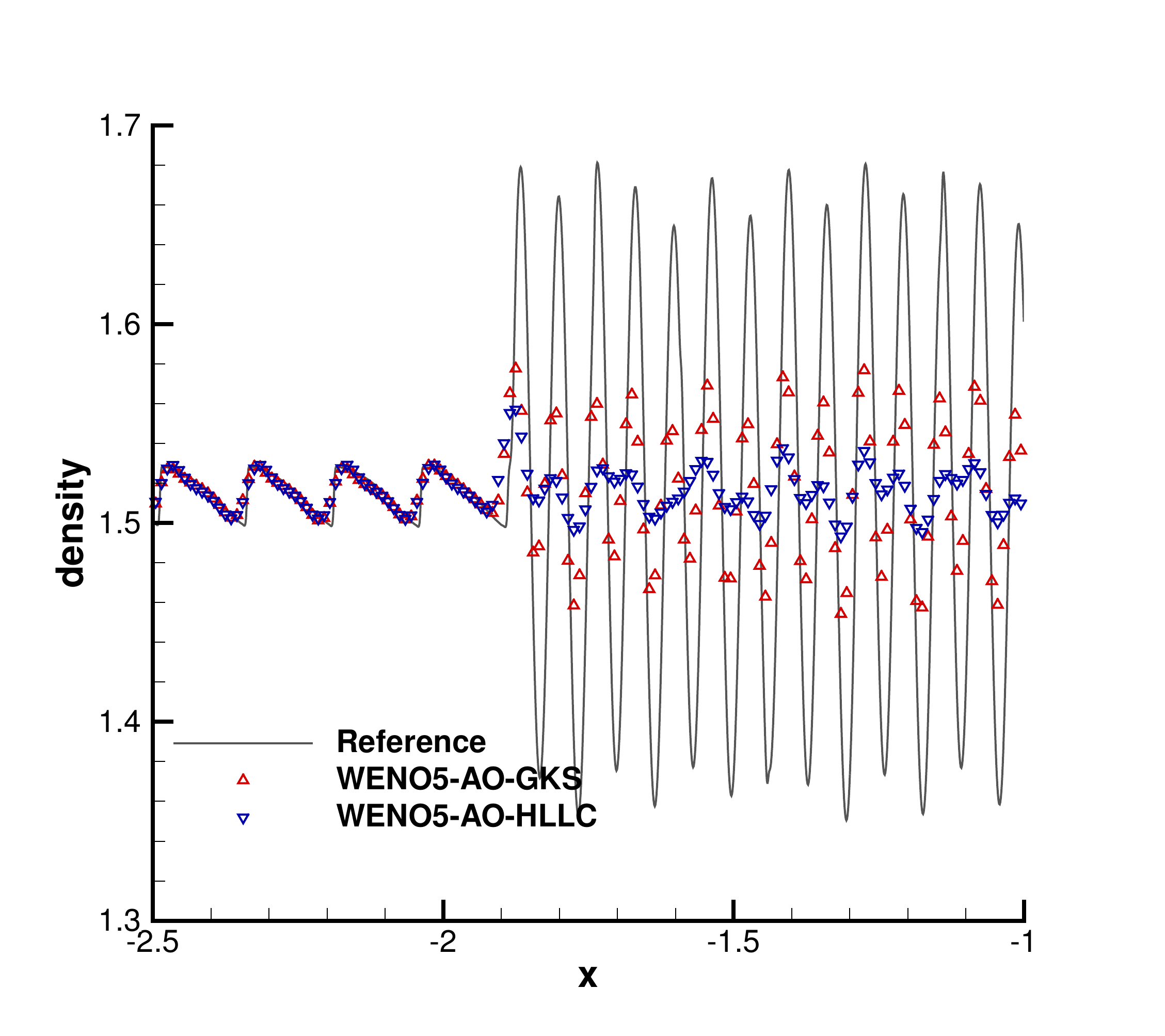}		
	}	
	\caption{Titarev-Toro problem by WENO5-AO-GKS scheme and WENO5-AO-HLLC scheme. Density distribution with mesh number 1000 at $t=5.0$. Left figure shows the whole domain; right figure shows the enlarged domain. The CFL number is 0.5.}
	\label{Titarey-Toro problem by GKS and HLLC with CFL 0.5}
\end{figure}

\begin{figure}[htbp]
	\centering
	\subfigure{
		\includegraphics[height=6cm]{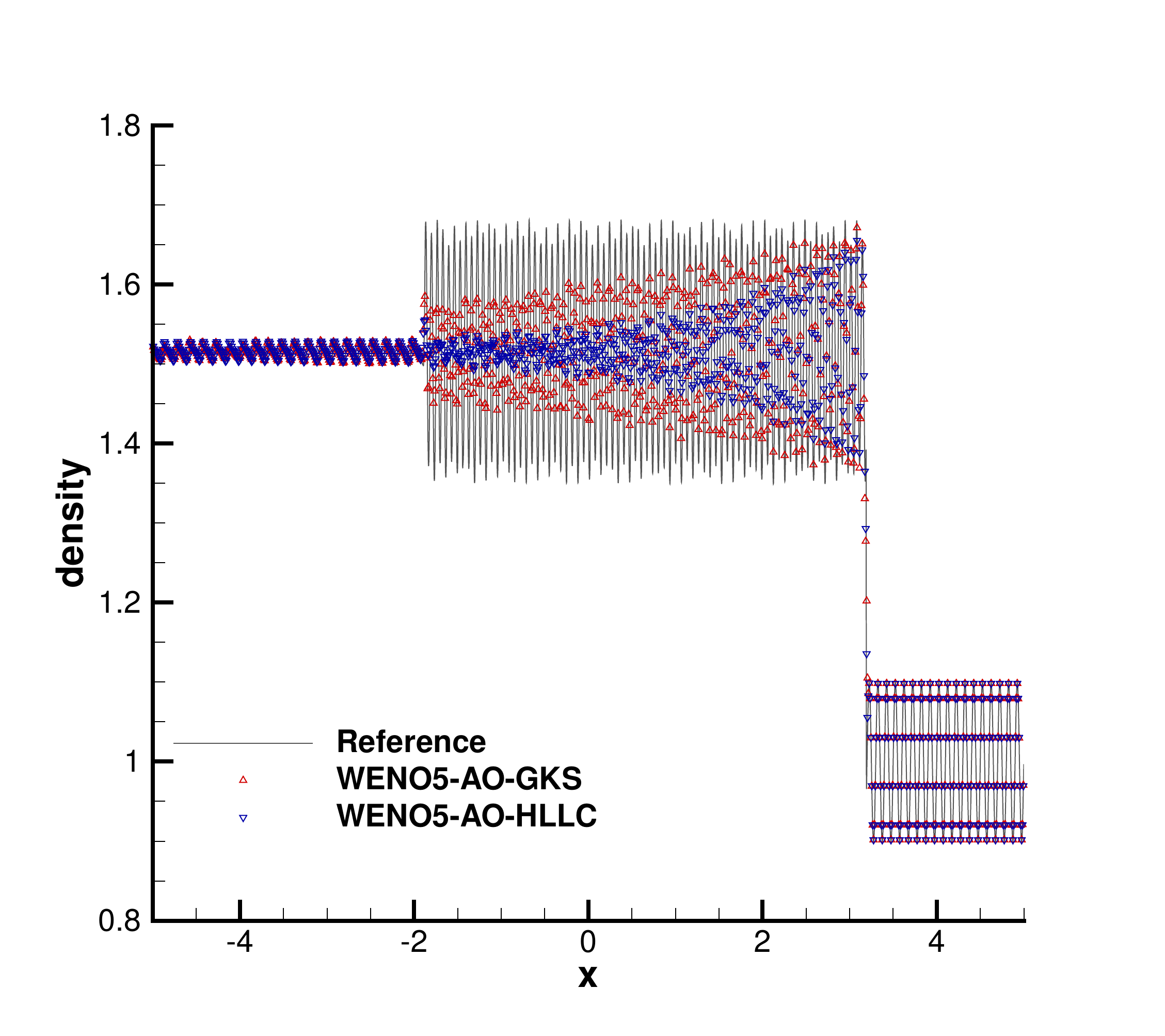}		
	}
	\quad
	\subfigure{
		\includegraphics[height=6cm]{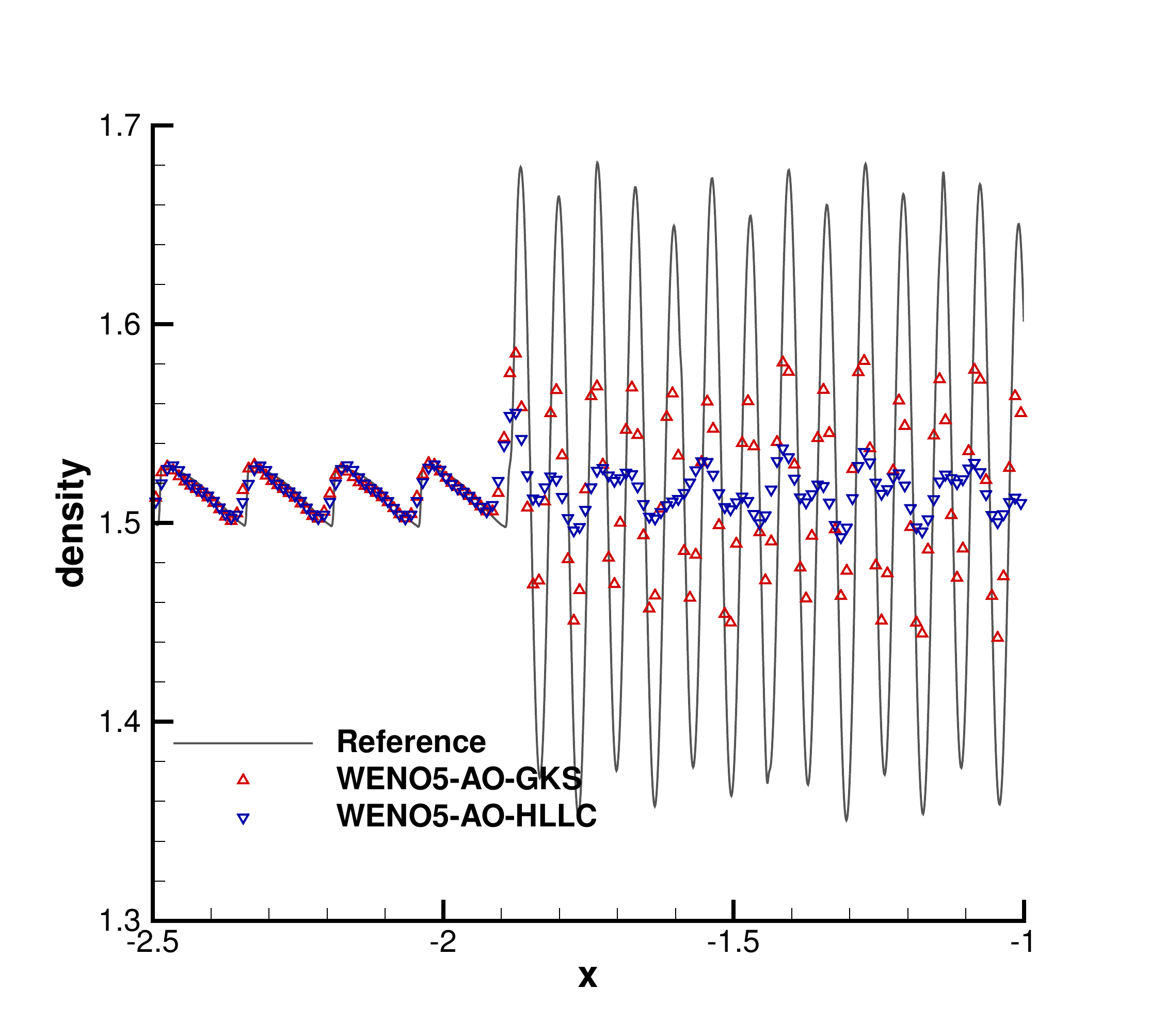}		
	}	
	\caption{Titarev-Toro problem by WENO5-AO-GKS scheme and WENO5-AO-HLLC scheme. Density distribution with mesh number 1000 at $t=5.0$. Left figure shows the whole domain; right figure shows the enlarged domain. The CFL number is 1.}
	\label{Titarey-Toro problem by GKS and HLLC with CFL 1.0}
\end{figure}

Titarev-Toro problem is an inviscid flow problem with a shock wave impinging into a high-frequency density perturbation \cite{Titarev-Toroproblem2004}. This problem consists of a main shock, a high gradient smooth post-shock region and multiple shocklets developed later. To represent these flow structures, a high-order scheme is needed. The initial condition is given by,
\begin{equation*}
(\rho,U,p)=\left\{\begin{aligned}
&(1.515695, 0.523346, 1.80500), & &-5.0 \le x \le -4.5,\\
&(1+0.1\text{sin}(20\pi x),0.0,1.0),   & &-4.5 < x \le 5.0.
\end{aligned} \right.
\end{equation*}	
The computational domain is $\left[-5,5\right]$ with a mesh of 1000 cells. Two CFL numbers, 0.5 and 1.0, are employed for both WENO5-AO-GKS and WENO5-AO-HLLC, and the results at the output time $t=5.0$ are presented in Figure \ref{Titarey-Toro problem by GKS and HLLC with CFL 0.5} and Figure \ref{Titarey-Toro problem by GKS and HLLC with CFL 1.0}, respectively. The results show that WENO5-AO-GKS scheme is more accurate than WENO5-AO-HLLC scheme at both CFL numbers, especially in the region behind the interaction of shock wave with the smooth acoustic wave.
These results may indicate the importance of time accurate flux in the simulation of high frequency unsteady flow.

\subsubsection{Le Blanc problem}
Le Blanc problem is a class of 1-D Riemann problems with initially high ratios for density and pressure \cite{LeBlancproblem}. Therefore, an extremely strong rarefaction wave is generated in the high-pressure region.
The initial condition here is chosen as,
\begin{equation*}
(\rho,U,p)=\left\{\begin{aligned}
&(10^M, 0, 10^M), & & 0\le x \le 0.3,\\
&(1,0,1),  & & 0.3 < x \le 1.
\end{aligned} \right.
\end{equation*}		

\begin{figure}[htbp]
	\centering
	\subfigure{
		\includegraphics[height=6cm]{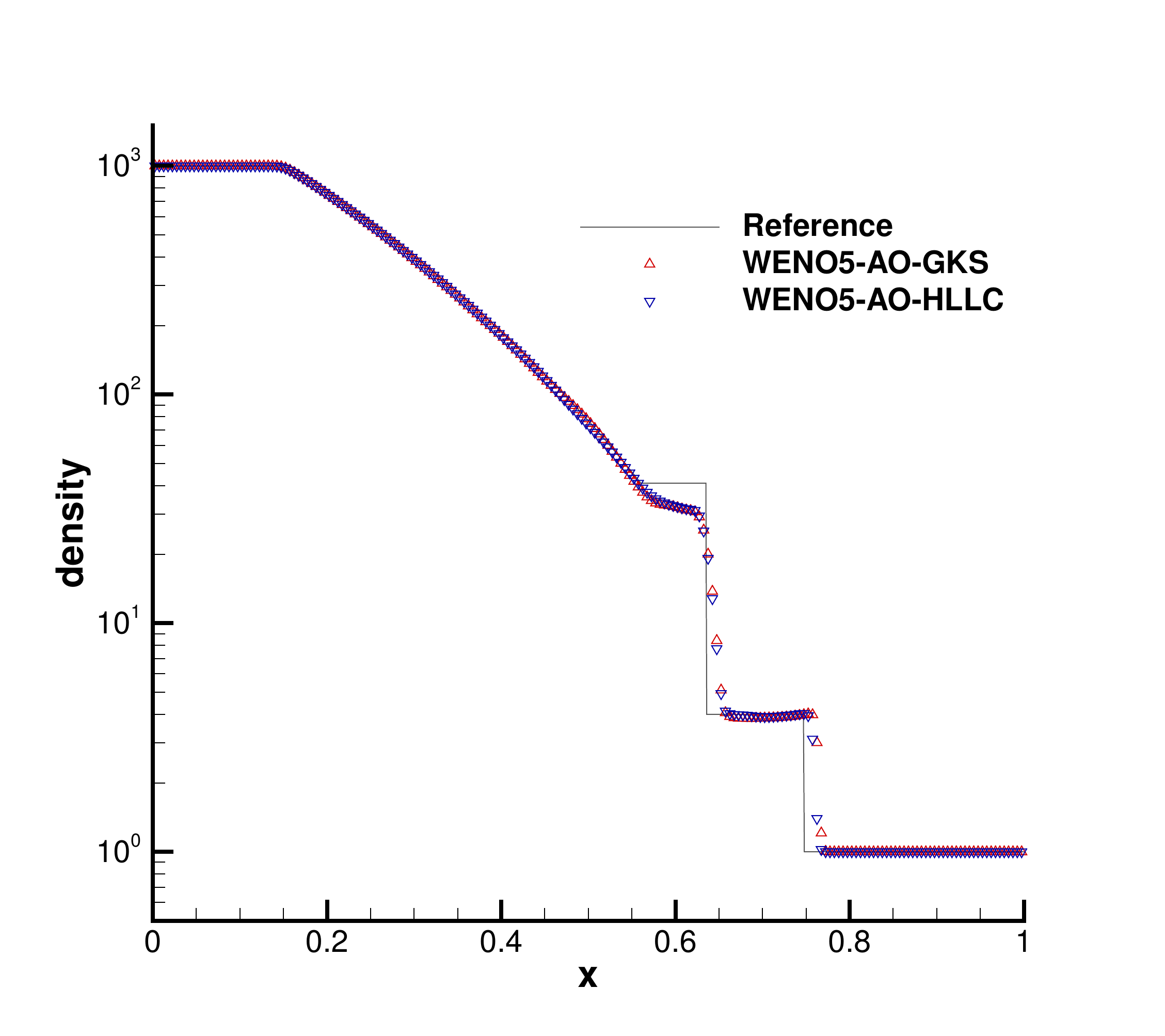}	
	}
	\quad
	\subfigure{
		\includegraphics[height=6cm]{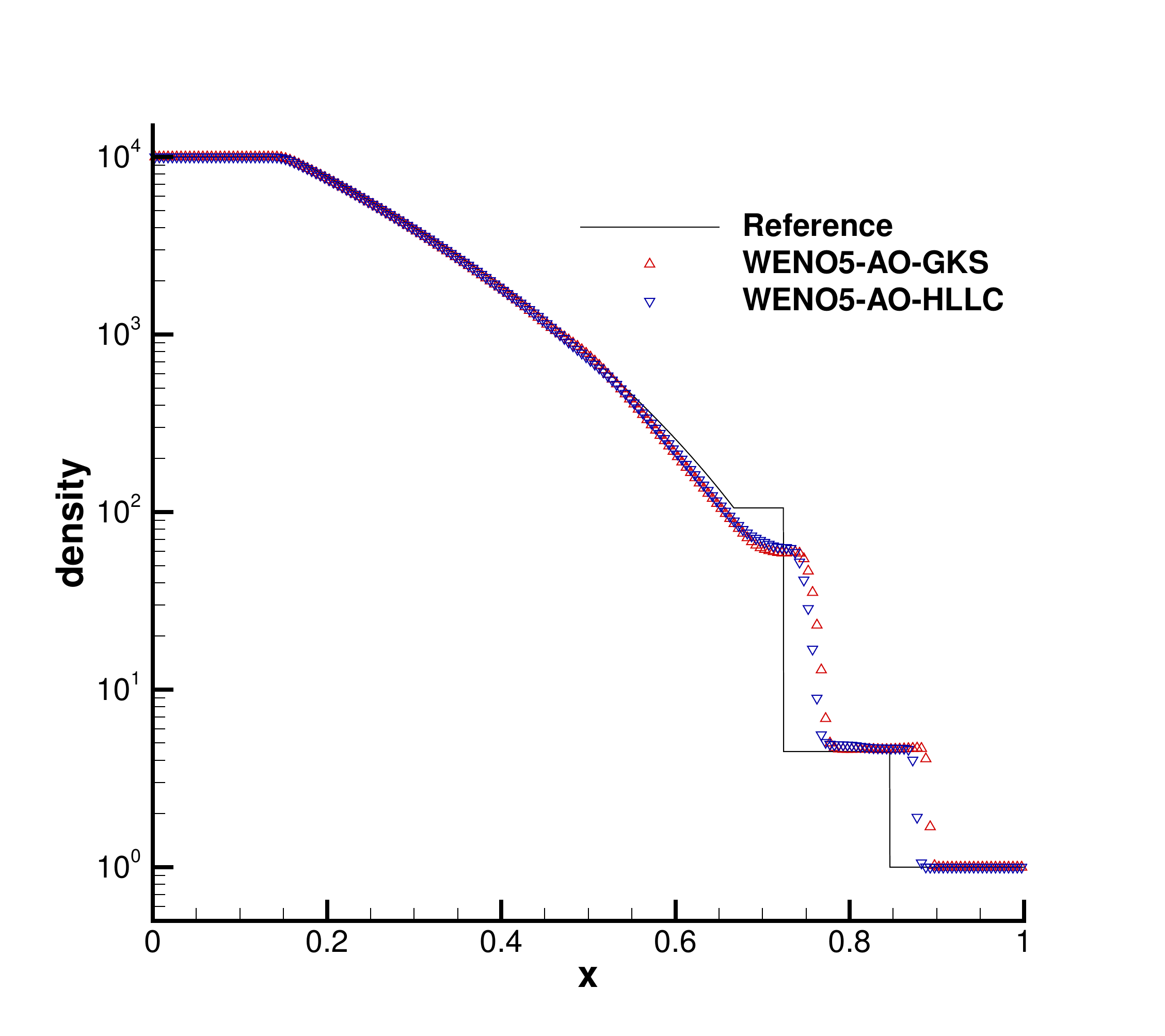}	
	}
	
	\subfigure{
		\includegraphics[height=6cm]{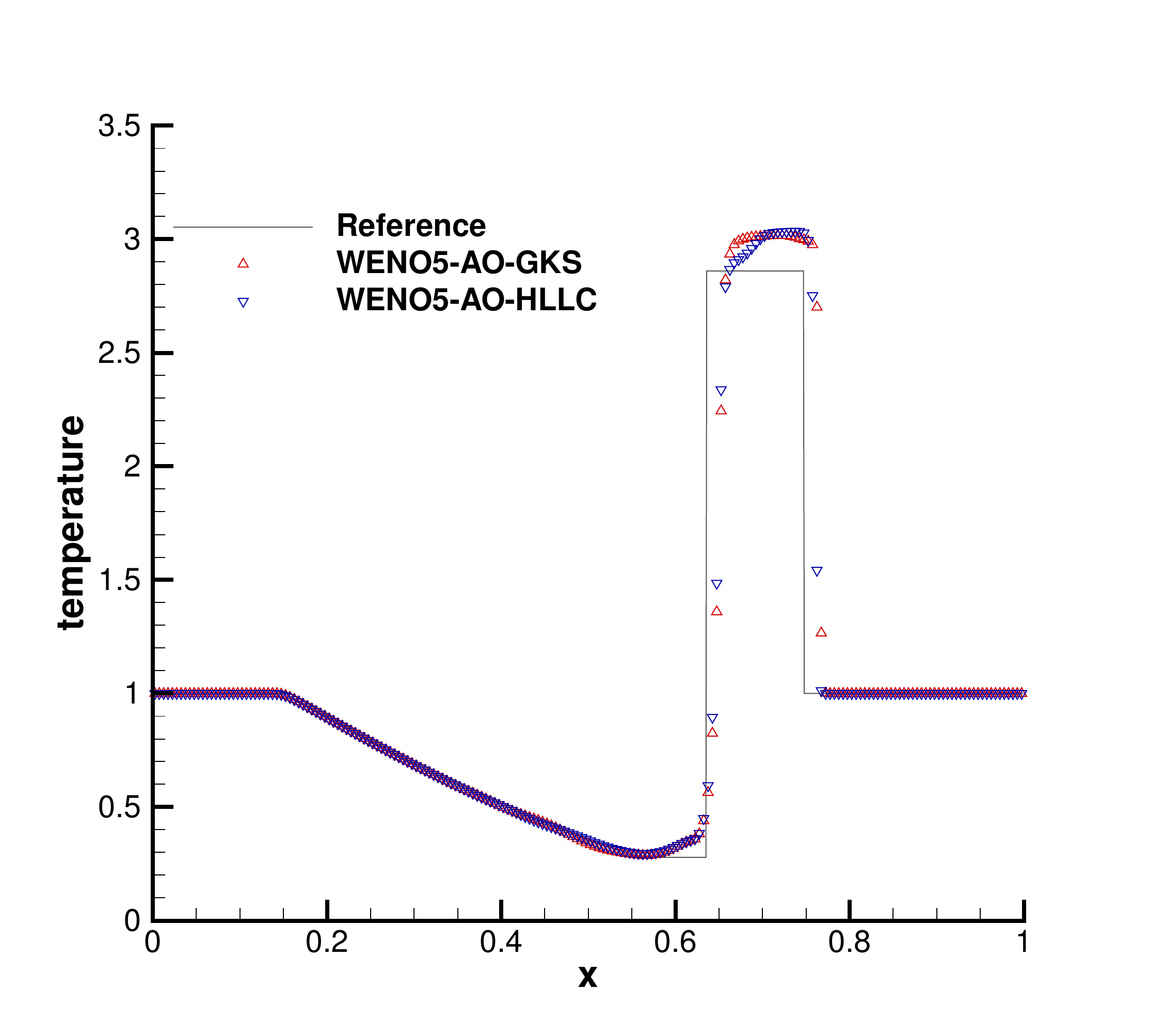}	
	}
	\quad
	\subfigure{
		\includegraphics[height=6cm]{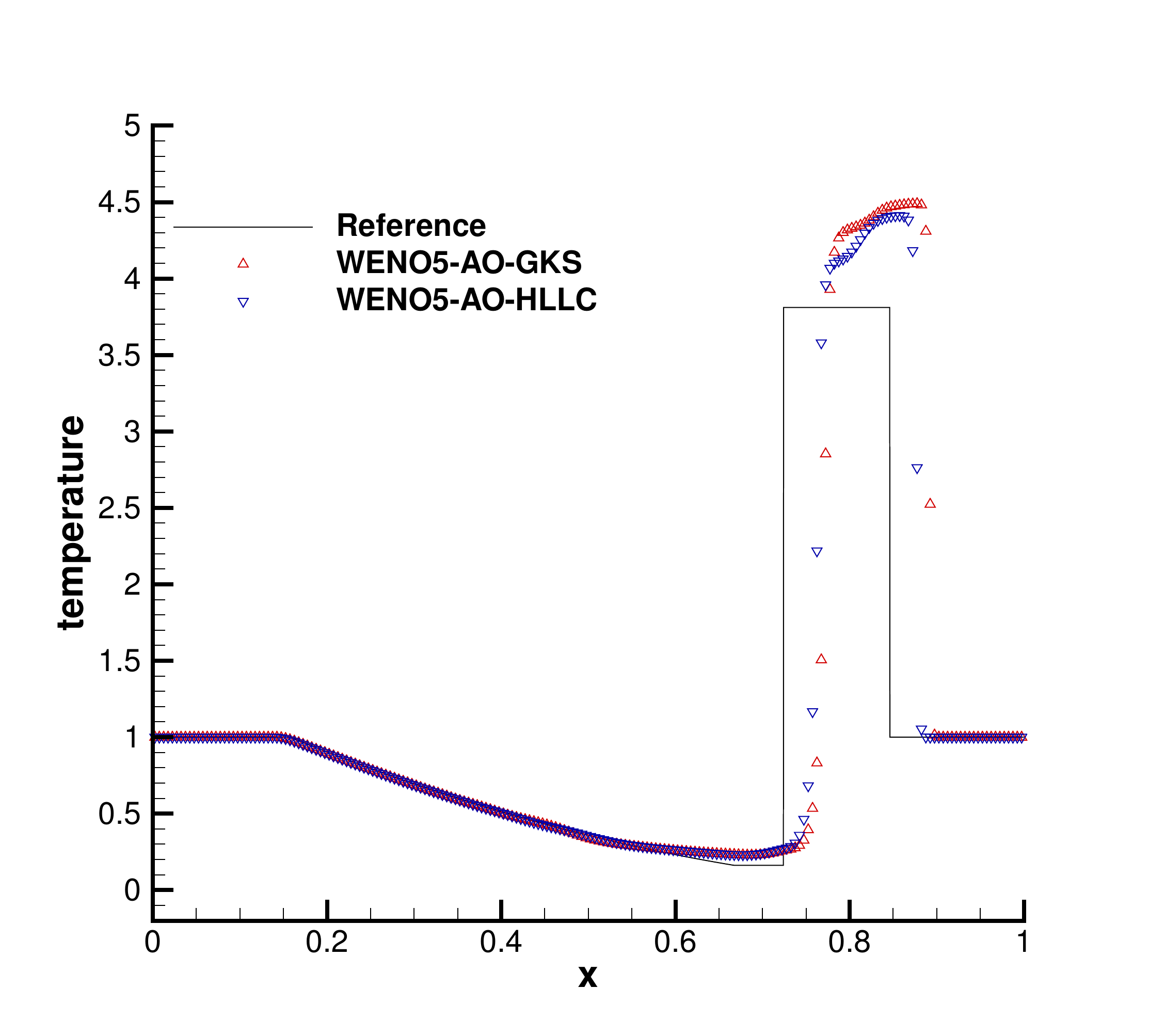}	
	}
	
	\subfigure{
		\includegraphics[height=6cm]{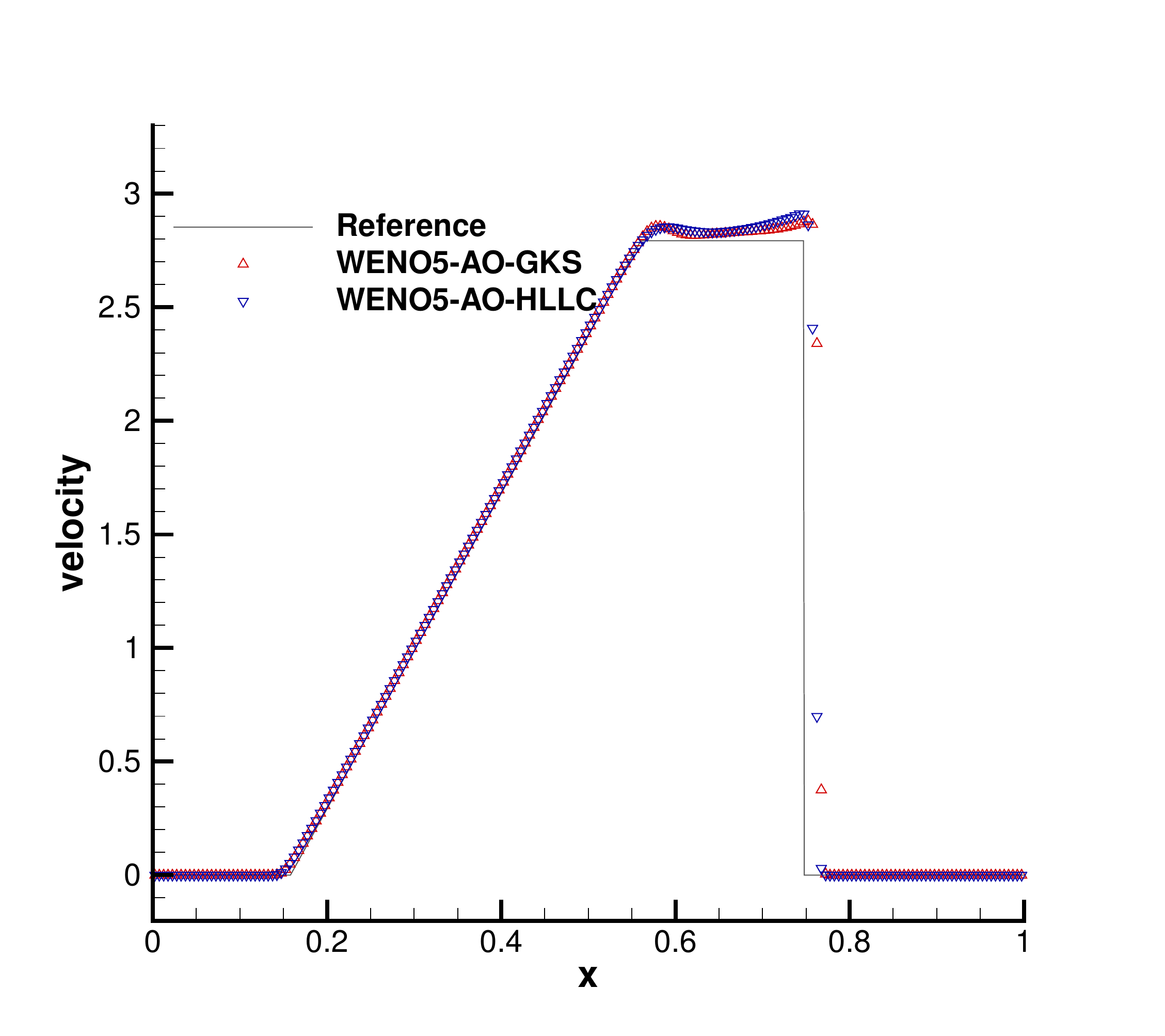}	
	}
	\quad
	\subfigure{
		\includegraphics[height=6cm]{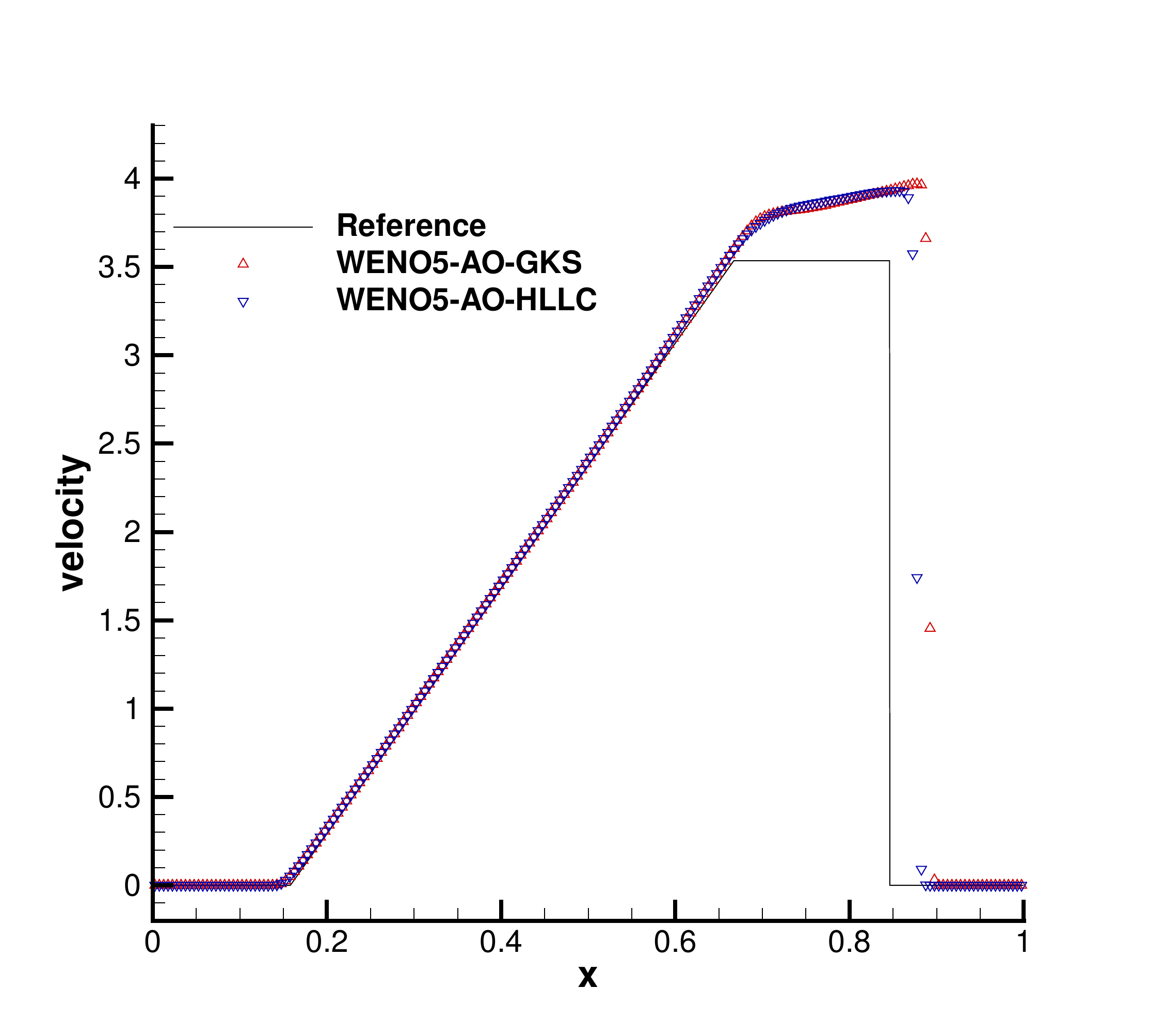}	
	}		
	\caption{Le Blanc problem with initial pressure ratio $10^3$ (left three figures) and $10^4$ (right three figures) by WENO5-AO-GKS scheme and WENO5-AO-HLLC scheme. For all figures, CFL number is 0.5, the mesh number is 200 and the output time is $t=0.12$.}
	\label{Le Blanc problem by GKS and HLLC}
\end{figure}

Here Le Blanc problem with initial pressure ratio $10^3$ and $10^4$ was calculated by WENO5-AO-GKS scheme and WENO5-AO-HLLC scheme, and the profiles of density, temperature, and pressure at $t=0.12$ are presented in Figure \ref{Le Blanc problem by GKS and HLLC}. For this case, CFL number is 0.5. For both two schemes, there exist discrepancy in the vicinity of the shock wave, which has also been observed in the previous research, especially in the coarse mesh case \cite{LeBlancproblem}. Both schemes present a similar performance in this case.

\subsubsection{Noh problem}
\begin{figure}[htbp]
	\centering
	\subfigure{
		\includegraphics[height=6cm]{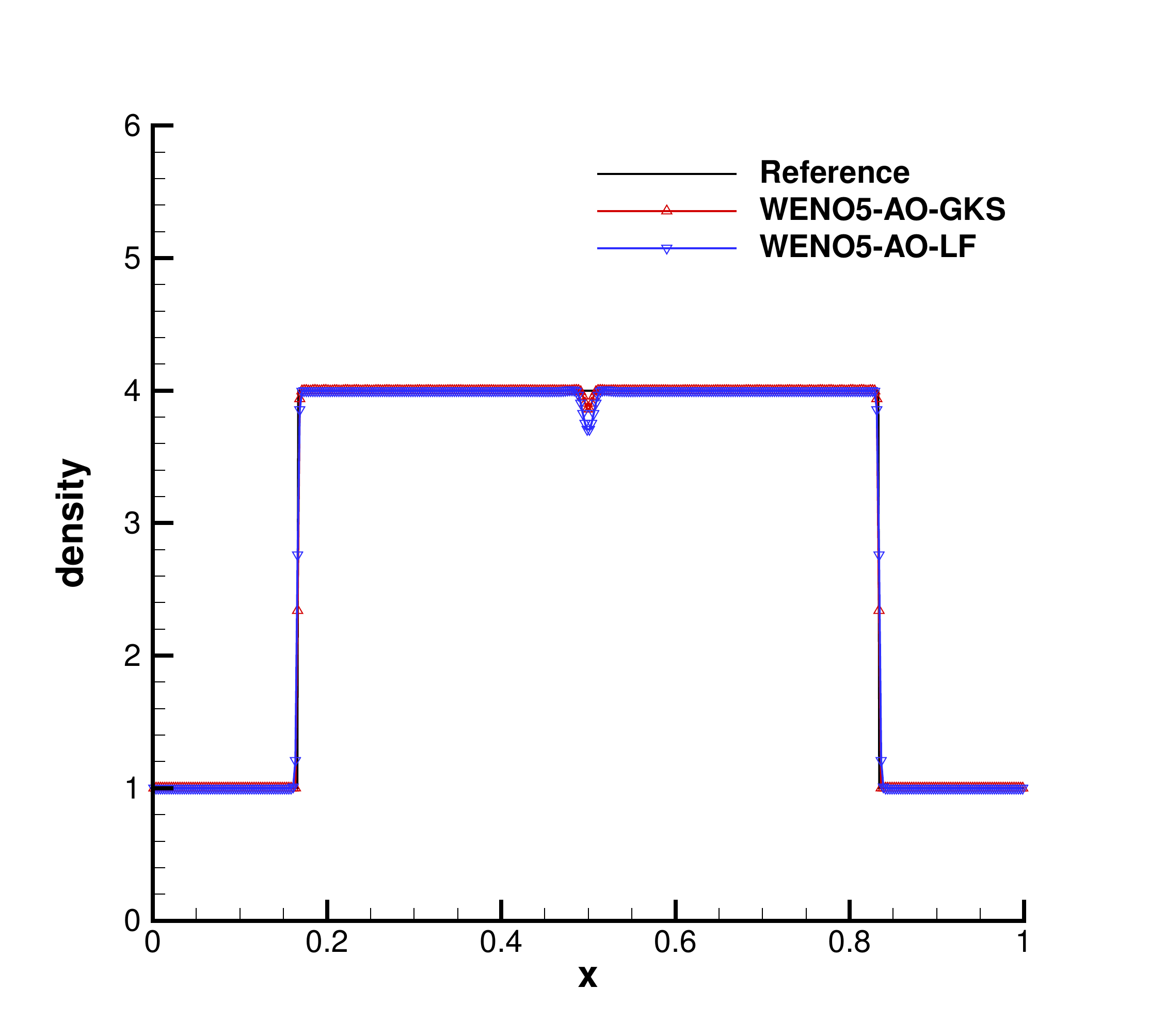}	
	}
	\quad
	\subfigure{
		\includegraphics[height=6cm]{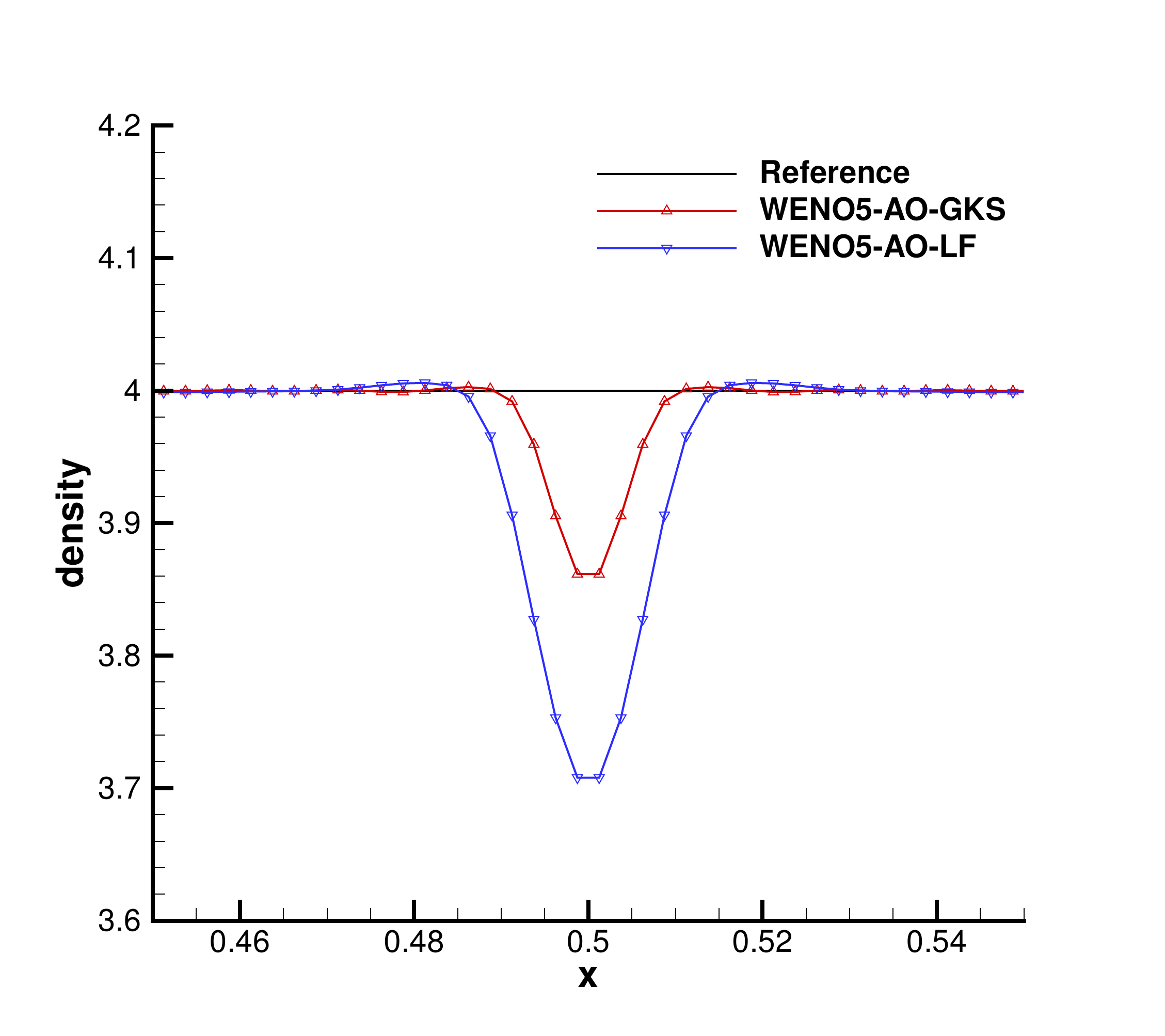}
	}
	
	\subfigure{
		\includegraphics[height=6cm]{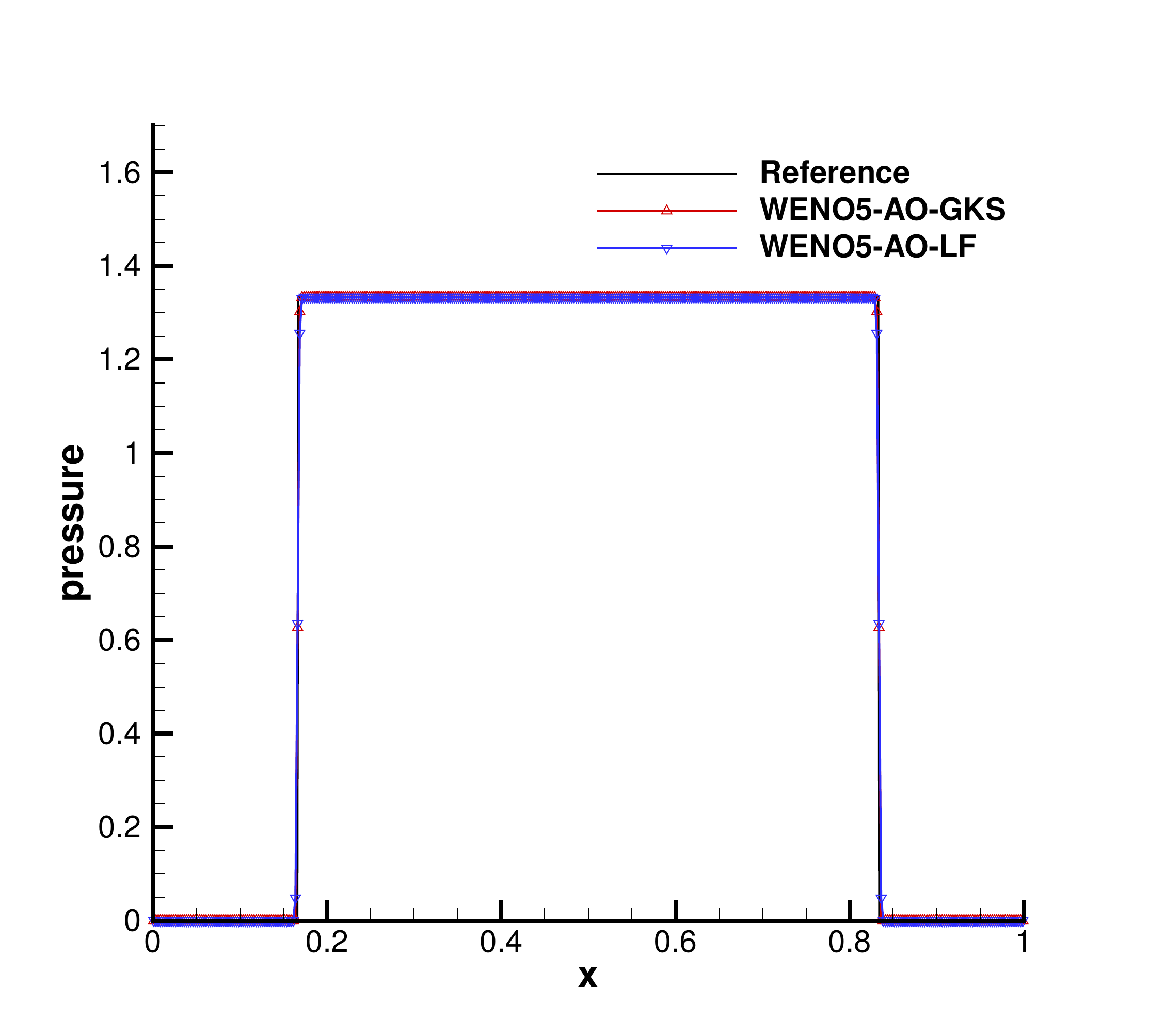}	
	}
	\quad
	\subfigure{
		\includegraphics[height=6cm]{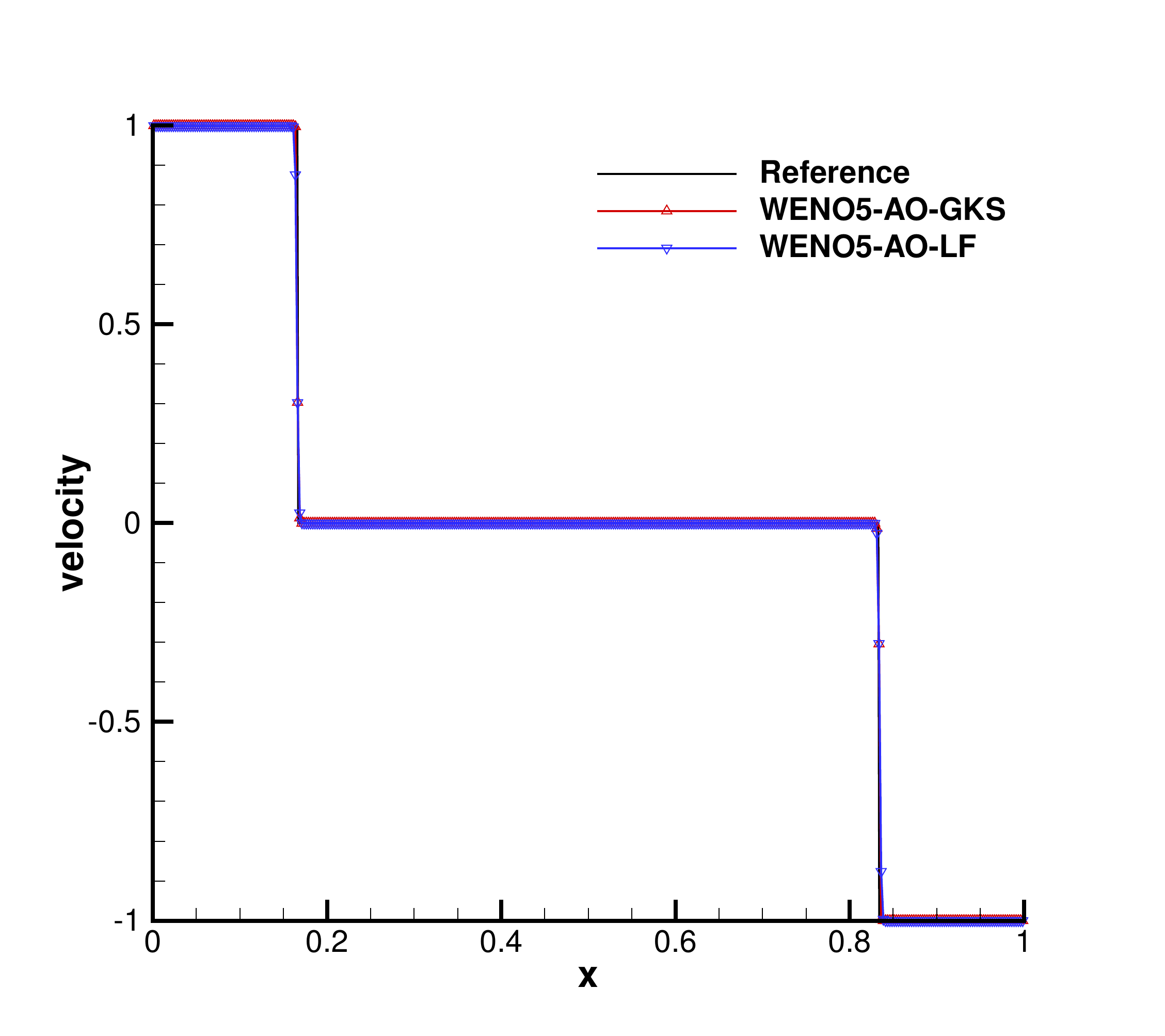}	
	}			
	\caption{Noh problem by WENO5-AO-GKS scheme and WENO5-AO-LF scheme. The density, pressure, and velocity profiles respectively with CFL number 0.5 are shown. For all figures, the mesh number is 400 and the output time is $t=1.0$. The WENO-AO-HLLC fails for this test case.}
	\label{Noh problem by GKS and LF}
\end{figure}

Noh problem consists of two strong shocks moving from center to left and right side respectively \cite{Nohproblem1987}.
The initial condition is as follows,
\begin{equation*}
(\rho,U,p)=\left\{\begin{aligned}
&(1,1,10^{-6}),   & & 0 \le x \le 0.5,\\
&(1,-1,10^{-6}),  & & 0.5 < x \le 1.
\end{aligned} \right.
\end{equation*}
The computational domain is $\left[0,1\right]$, which is covered by 400 cells. In this problem, the specific heat ratio is $\gamma=5/3$. The output time is $t=1.0$. The results are presented in Figure \ref{Noh problem by GKS and LF}. It is worth noting that WENO5-AO-HLLC scheme blows up for this problem while WENO5-AO-GKS scheme can work well. Besides, as a comparison, the WENO5-AO-LF scheme is adopted for this problem. The WENO5-AO-LF scheme means that, only HLLC solver in WENO5-AO-HLLC scheme is replaced by Lax-Friedrich solver. The results show that both two schemes can resolve the shock very well, although the density profiles exist a weak dip at the central region for both schemes. Besides, the result of WENO5-AO-GKS scheme shows a weaker dip.

\subsection{2-D tests}

\subsubsection{Forward step problem}
\begin{figure}[htbp]
	\centering
	\subfigure{
		\includegraphics[height=2.55cm]{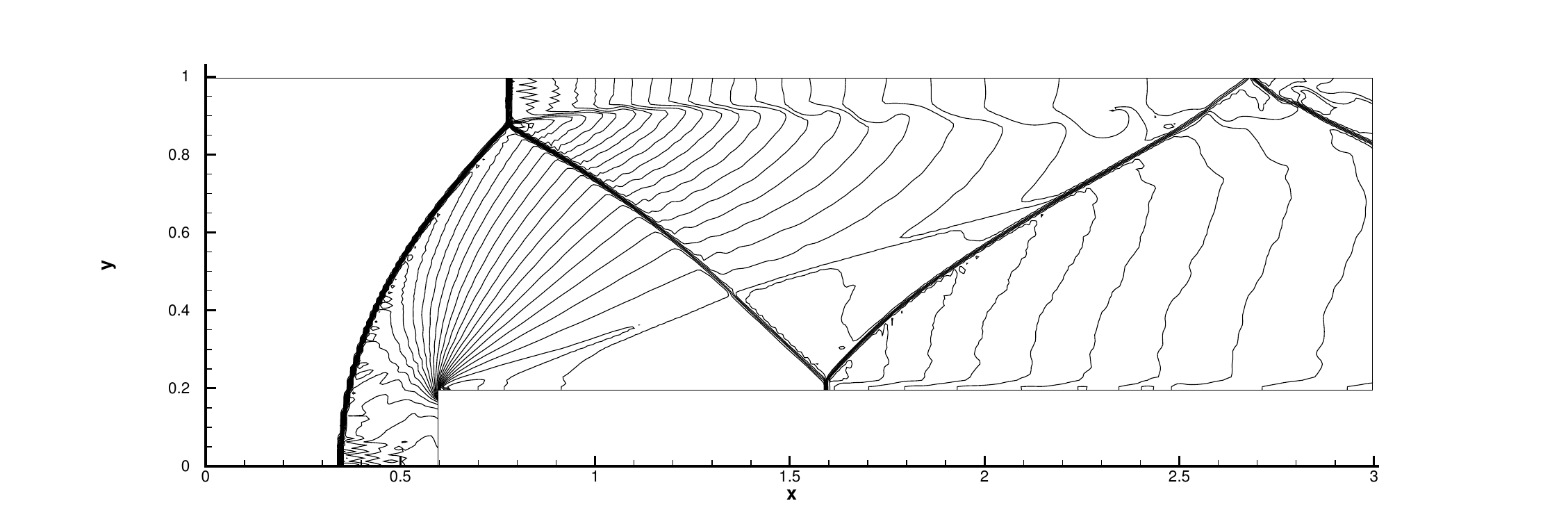}	
	}
	\quad
	\subfigure{
		\includegraphics[height=2.55cm]{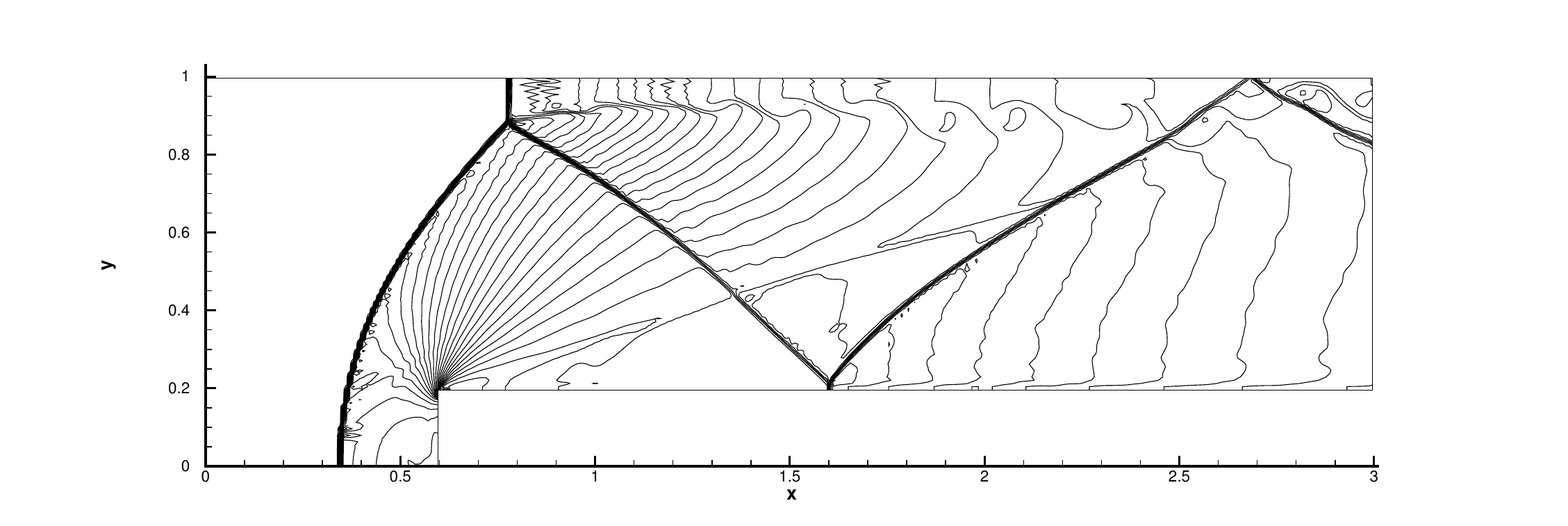}
	}
	
	\subfigure{
		\includegraphics[height=2.55cm]{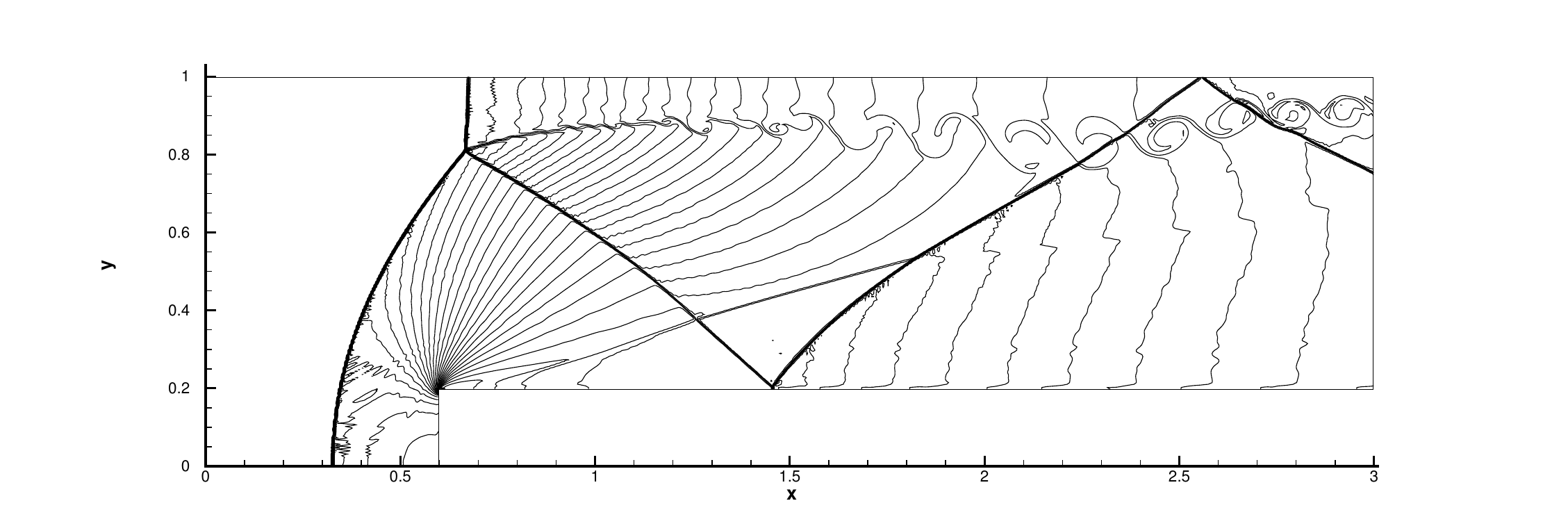}	
	}
	\quad
	\subfigure{
		\includegraphics[height=2.55cm]{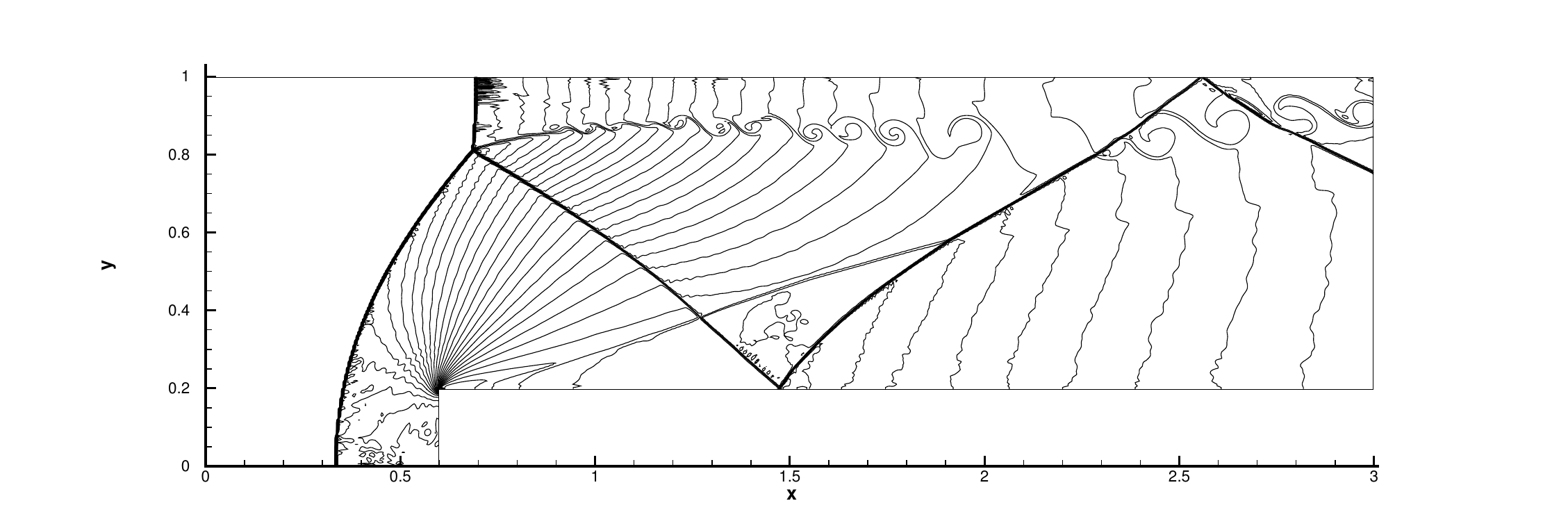}
	}
	
	\subfigure{
		\includegraphics[height=2.55cm]{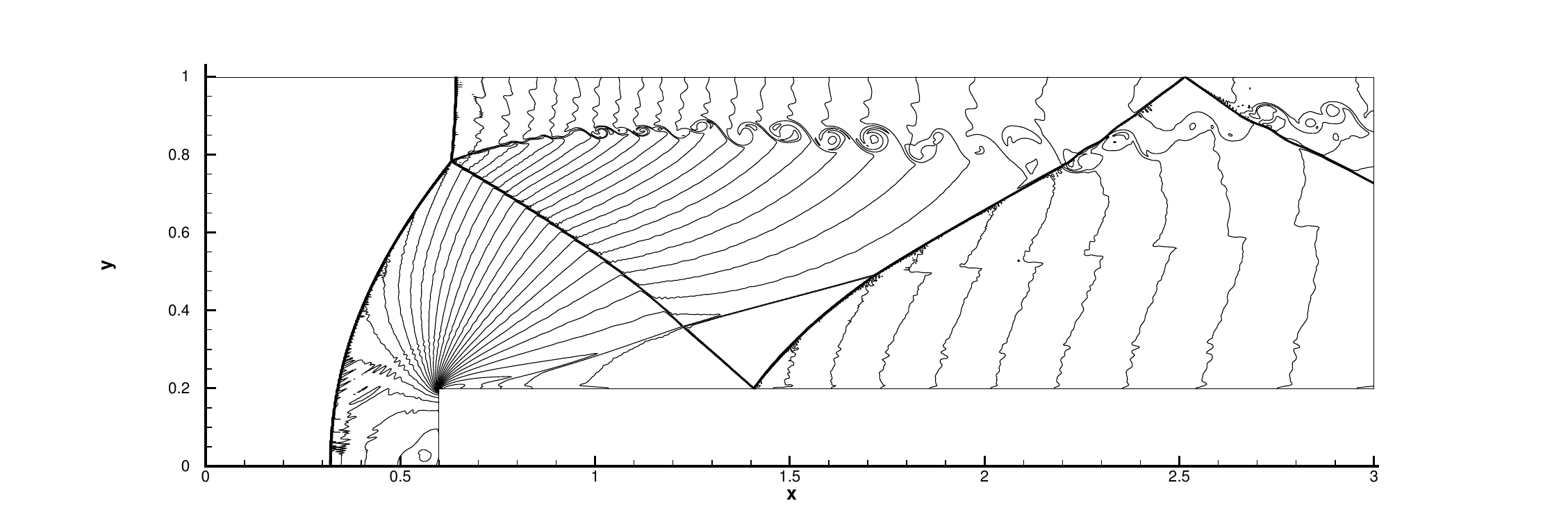}	
	}
	\quad
	\subfigure{
		\includegraphics[height=2.55cm]{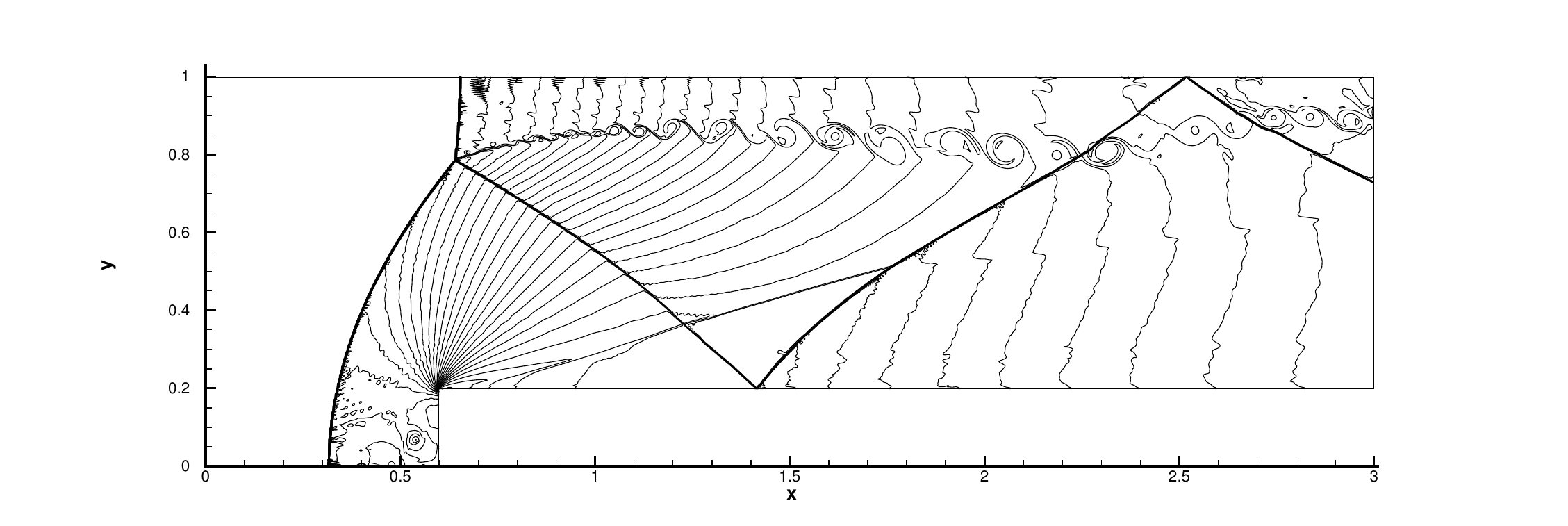}
	}
	\caption{Mach 3 forward step problem by WENO5-AO-GKS scheme (left) and WENO5-AO-HLLC scheme (right). Density distribution with different mesh size at $t=4.0$. The mesh size of the top figure, middle figure, and bottom figure are 1/120, 1/240, and 1/360. The CFL number is 0.8 for both WENO-AO-GKS and WENO5-AO-HLLC. 30 equally spaced contours from 0.2 to 4.7 are plotted.}
	\label{Mach 3 Forward step problem by GKS and HLLC with CFL 0.8}
\end{figure}

The forward step problem proposed by Woodward and Colella \cite{Forwardstepproblem1984} is an inviscid test case. A uniform flow with $Ma=3$ blows towards a wind tunnel containing a step. This wind tunnel size, is $[0,3]\times[0,1]$; the step is located at 0.6 from the left and has 0.2 high. The initial condition can be described as,
\begin{equation*}
(\rho,U,V,p)=(1,3,0,1/\gamma),
\end{equation*}
where $\gamma=1.4$.
The supersonic inlet and outlet boundary condition is employed for the left and right boundary respectively, while other boundaries are set as reflective boundary conditions.
It is worth remarking that, the ghost cells near the corner of the step $[0.6,0.2]$ are given as follows: velocity $U$ is given by the value obtained through applying the reflective boundary condition for the upper flow region; velocity $V$ is given by the value obtained through applying reflective boundary condition for the left flow region; density $\rho$ and pressure $p$ are given by algebraically averaging the corresponding values obtained through the reflective boundary condition for the upper and left flow regions.
The CFL number $0.8$ is used. The results are shown in Figure \ref{Mach 3 Forward step problem by GKS and HLLC with CFL 0.8}, respectively. For each case, three values of the mesh size, $\Delta x=\Delta y=1/120, 1/240, 1/360$, are taken, and the output time is $t=4.0$. The results show that both WENO5-AO-GKS scheme and WENO5-AO-HLLC scheme perform well when adopting a fine mesh. In the top region, both the triple-point structure and the vortex sheet can be captured clearly.

\subsubsection{Laminar boundary layer}	
\begin{figure}[htbp]
	\centering
	\subfigure{
		\includegraphics[height=8cm]{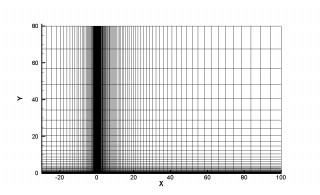}}		
	\caption{Mesh with $120\times32$ cells for laminar boundary layer case.}
	\label{Laminar boundary layer mesh}
\end{figure}

Laminar boundary layer is a standard test case for viscous flow \cite{GKS-lecture}. A plane with the characteristic length $L=100$ is placed from 0 to 100. The computation domain is $[-30,100]\times[0,80]$. Non-uniform mesh is adopted, which is shown in Figure \ref{Laminar boundary layer mesh}. The mesh number is $120\times32$. At the start point of plane, the minimal cell mesh $\Delta x$ and $\Delta y$ are 0.1 and 0.12 separately. The inlet flow is described by,
\begin{equation*}
(\rho,U,V,p)=(1,0.15,0,1/\gamma),
\end{equation*}
where $\gamma=1.4$.
In the case, kinematic viscosity coefficient is $\nu=1.5\times10^{-4}$, and thus $Re=U_\infty L/\nu=1.0\times10^5$ and $Ma=0.15$.
Besides, the adiabatic non-slip boundary condition is adopted on the plate, while the symmetric slip boundary condition is used for the bottom boundary of $[-30,0]$.
The outflow boundary condition is given at the right boundary.
The non-reflecting boundary condition is imposed on other boundaries.

\begin{figure}[htbp]
	\centering
	\subfigure{
		\includegraphics[height=6cm]{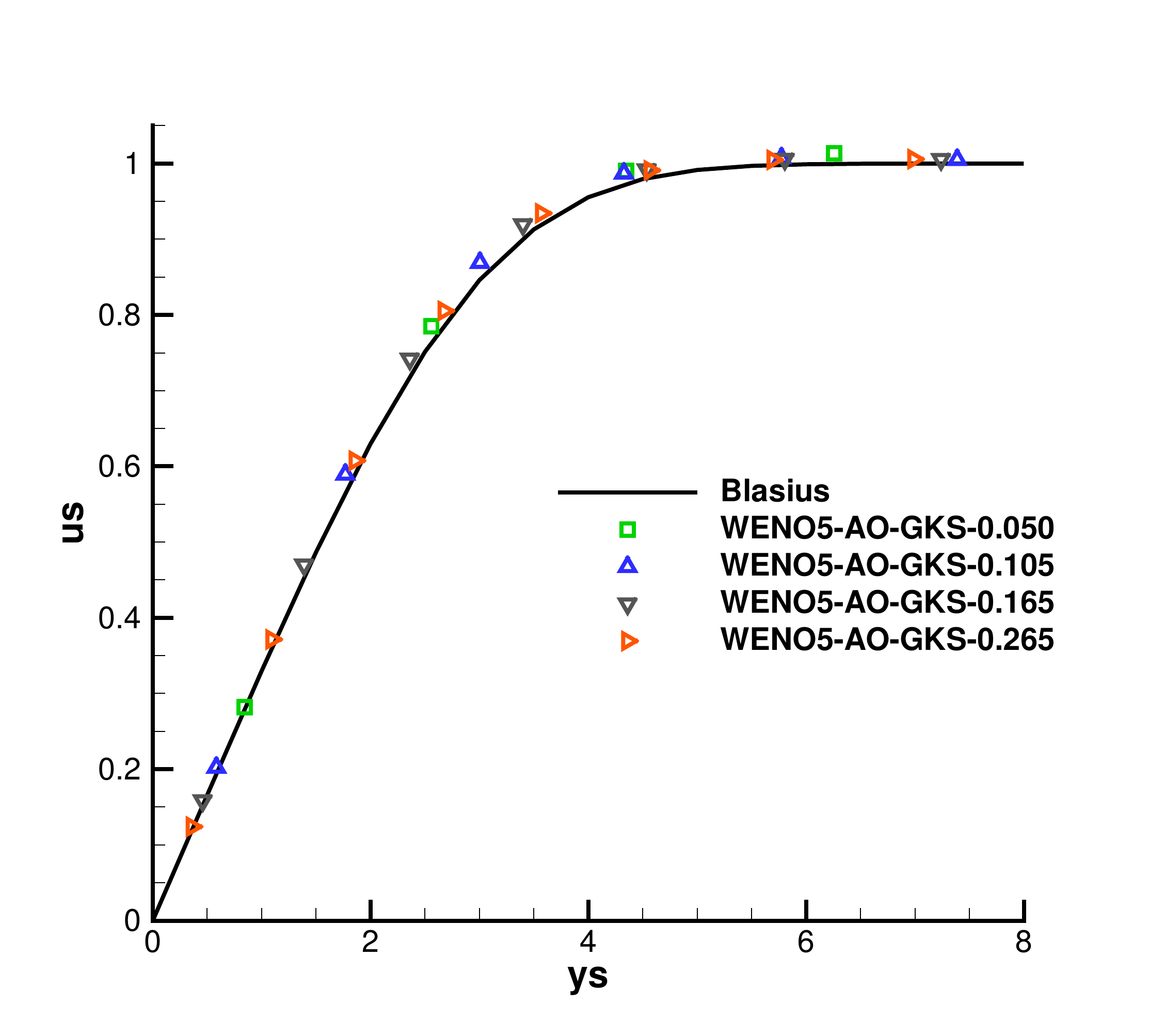}	
	}
	\quad
	\subfigure{
		\includegraphics[height=6cm]{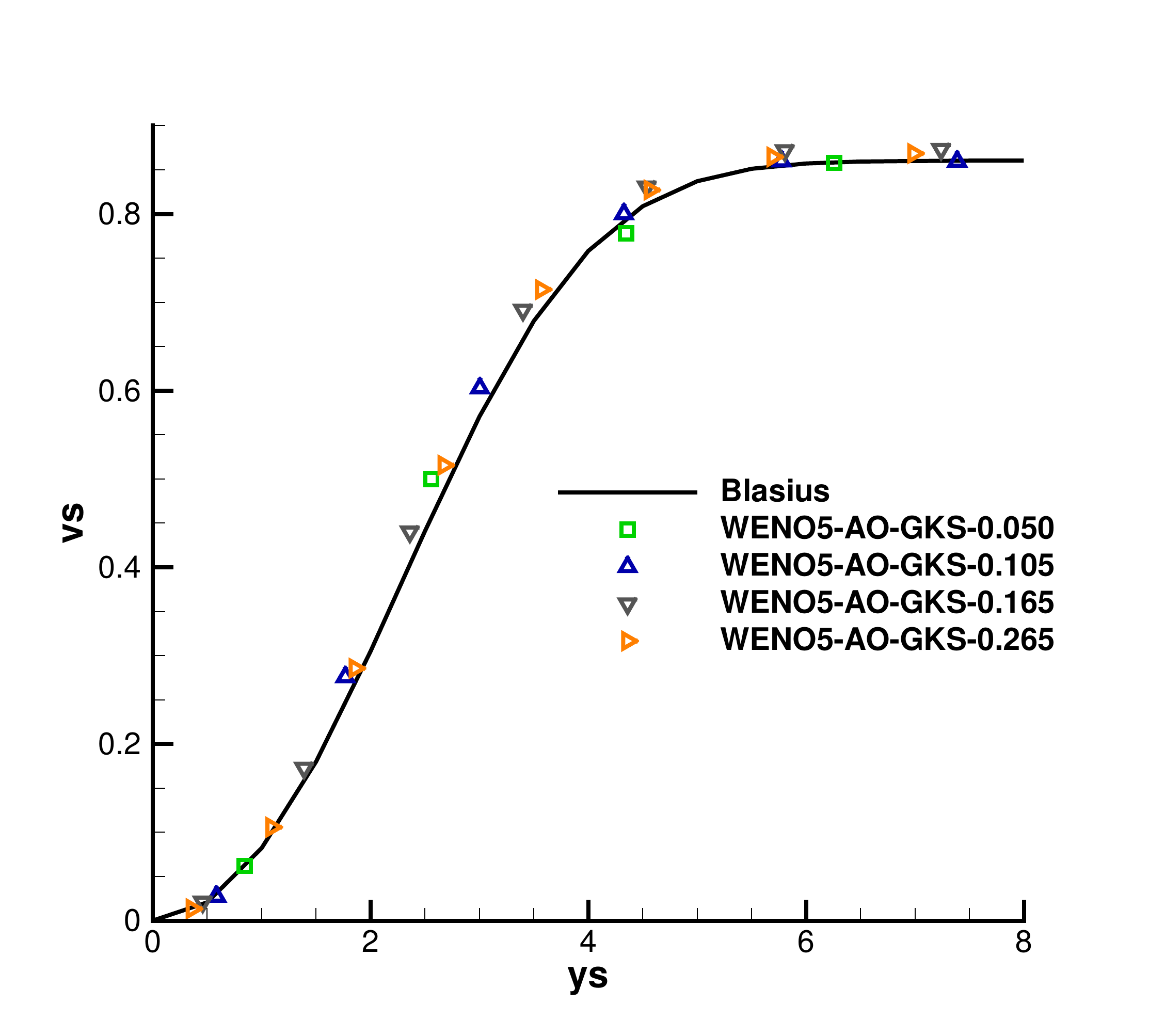}
	}		
	
	\subfigure{
		\includegraphics[height=6cm]{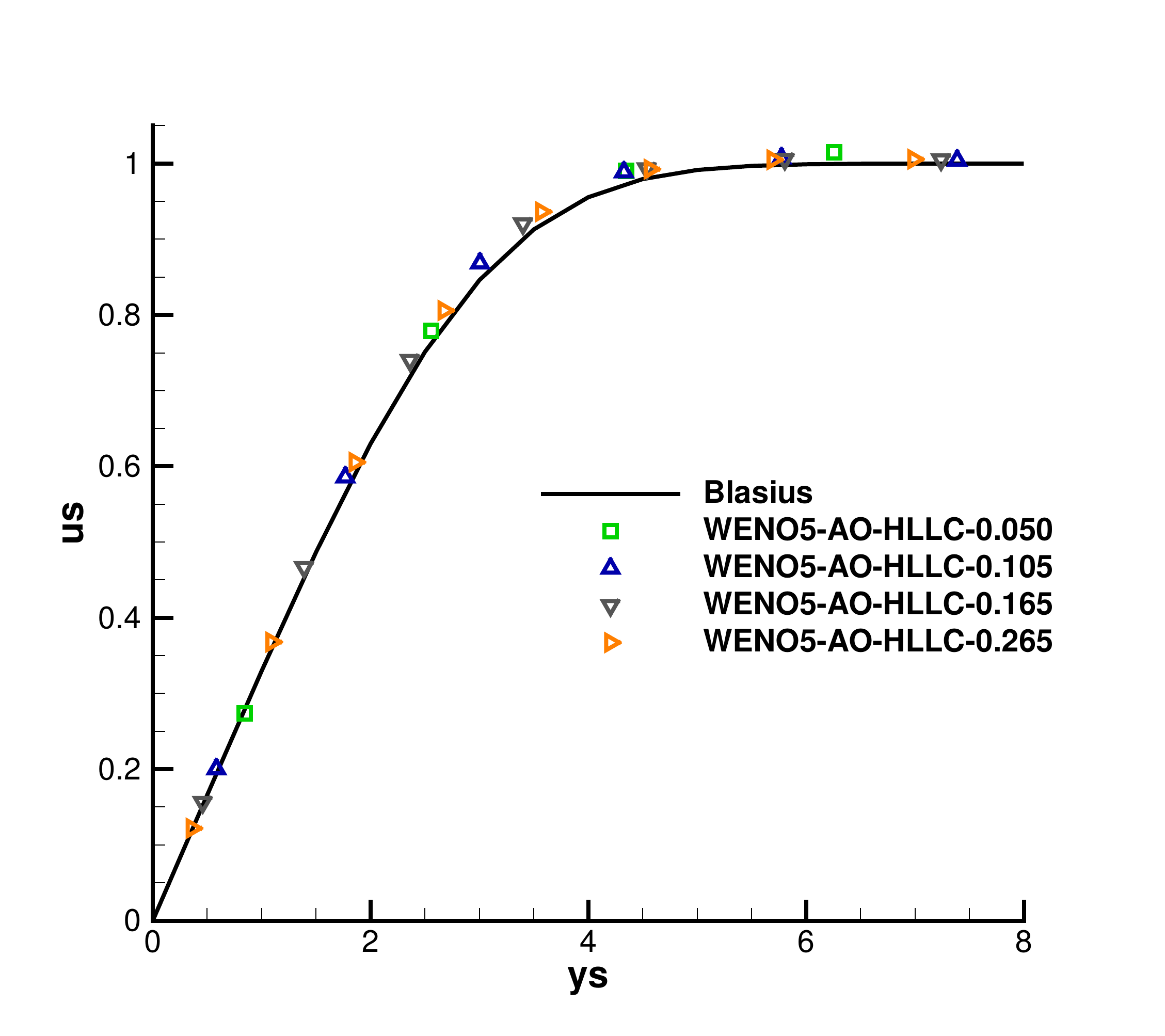}	
	}
	\quad
	\subfigure{
		\includegraphics[height=6cm]{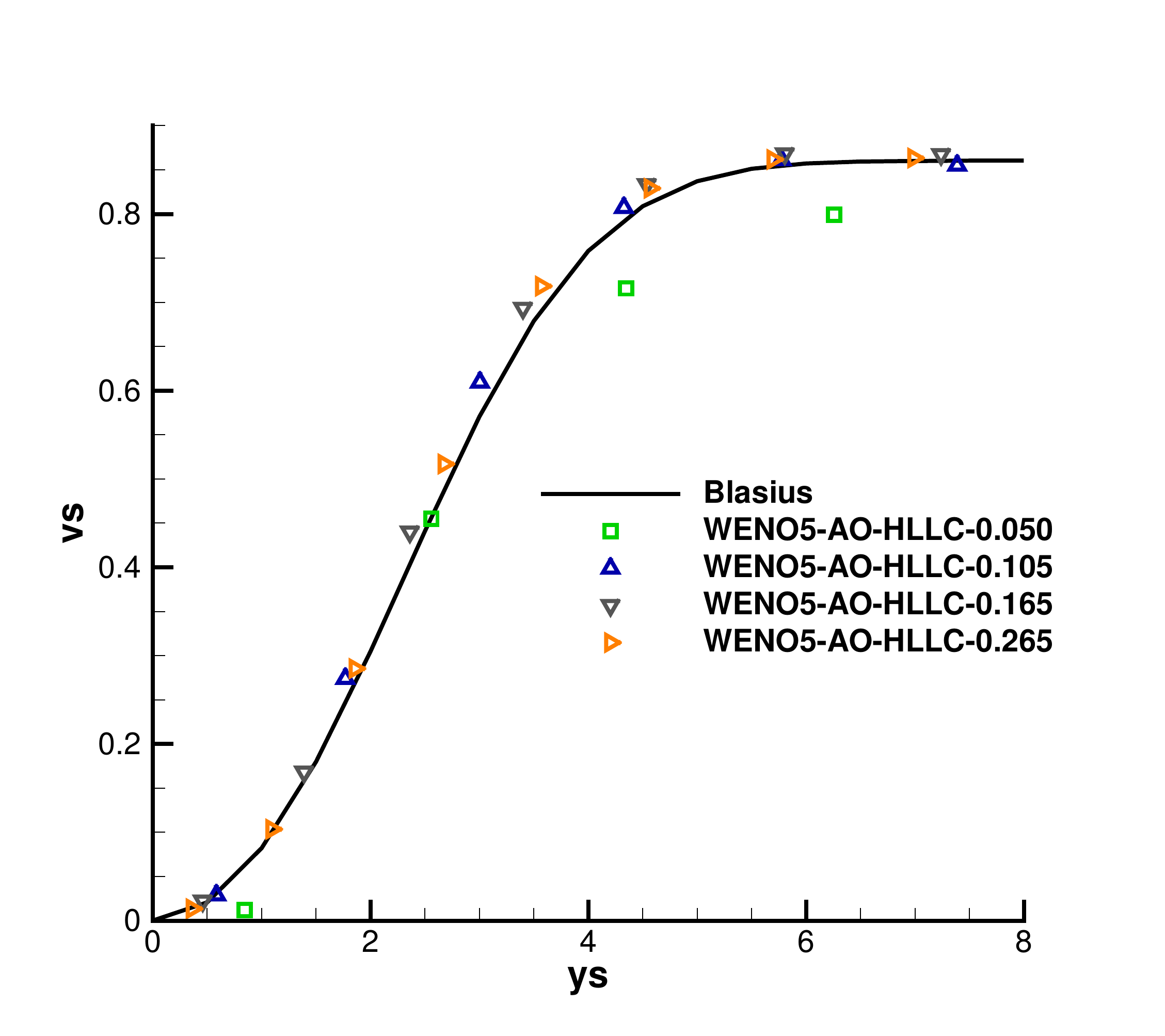}	
	}		
	\caption{Laminar boundary layer by WENO5-AO-GKS scheme (top two) and WENO5-AO-HLLC scheme (bottom two). For all figures, $Re = 1.0 \times 10^5$, $Ma = 0.15$, CFL number is 0.5, and the mesh number is $120\times32$.}
	\label{Laminar boundary layer problem by GKS and HLLC Re 100000}
\end{figure}

The results are presented in Figure \ref{Laminar boundary layer problem by GKS and HLLC Re 100000}, where the non-dimensional length $ys=y\sqrt{Re}/x$, and the non-dimensional velocity $us=U/U_\infty$, $vs = V\sqrt{Re_x}/U_\infty$, respectively. The values in the legend represent the location $x/L$.
From the results, both WENO5-AO-GKS and WENO5-AO-HLLC  are capable of capturing the velocity profile well in the boundary layer with several mesh cells. Close to the leading edge, WENO5-AO-GKS gives a slightly better $vs$ solution than WENO5-AO-HLLC at the location $x/L=0.050$.

\subsubsection{Double shear layer}
Double shear layer is a viscous problem involving a pair of doubly-periodic shear layers \cite{Doubleshearflow1995}. When the numerical method is not enough to resolve the flow field, non-physical vortexes will appear in the evolution stage. The ``thin" shear layer problem is studied in \cite{Doubleshearflow1995}, and the initial $U$ velocity is given by,
\begin{equation*}
U =\left\{\begin{aligned}
&\text{tanh}\left(k\left(y-0.25\right)\right), & & 0\le y \le 0.5,\\
&\text{tanh}\left(k\left(0.75-y\right)\right),  & & 0.5 < y \le 1,
\end{aligned} \right.
\end{equation*}
and the initial $V$ velocity, density, and pressure are given as follows,
\begin{equation*}
V = \delta \text{sin}\left(2 \pi x\right),~~
\rho = 1,~~
p = \frac{\rho U^2}{Ma^2\gamma},
\end{equation*}
where the shear layer width parameter $k=100$, the perturbation size $\delta=0.05$, the Mach number $Ma=0.15$, and the specific heat ratio $\gamma=1.4$. Besides, the kinetic viscosity is $\nu = 5.0 \times10^{-5}$. Periodic boundary condition is employed for all boundaries. The computational domain is $[0,1]\times[0,1]$, and mesh number is $256\times256$ in this case. Linear reconstruction is employed for both schemes in this test.

\begin{figure}[htbp]
	\centering
	\subfigure{
		\includegraphics[height=6cm]{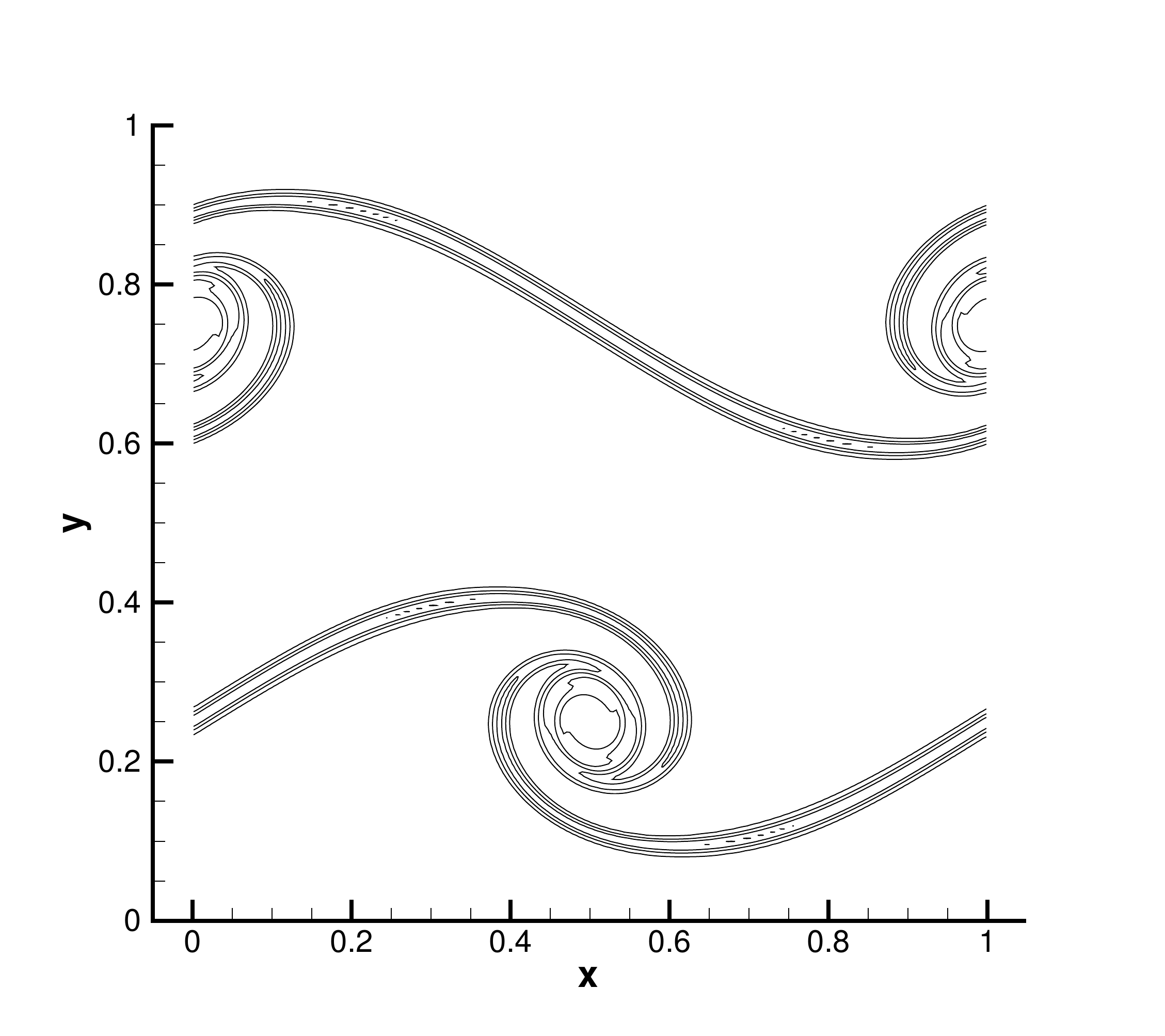}	
	}
	\quad
	\subfigure{
		\includegraphics[height=6cm]{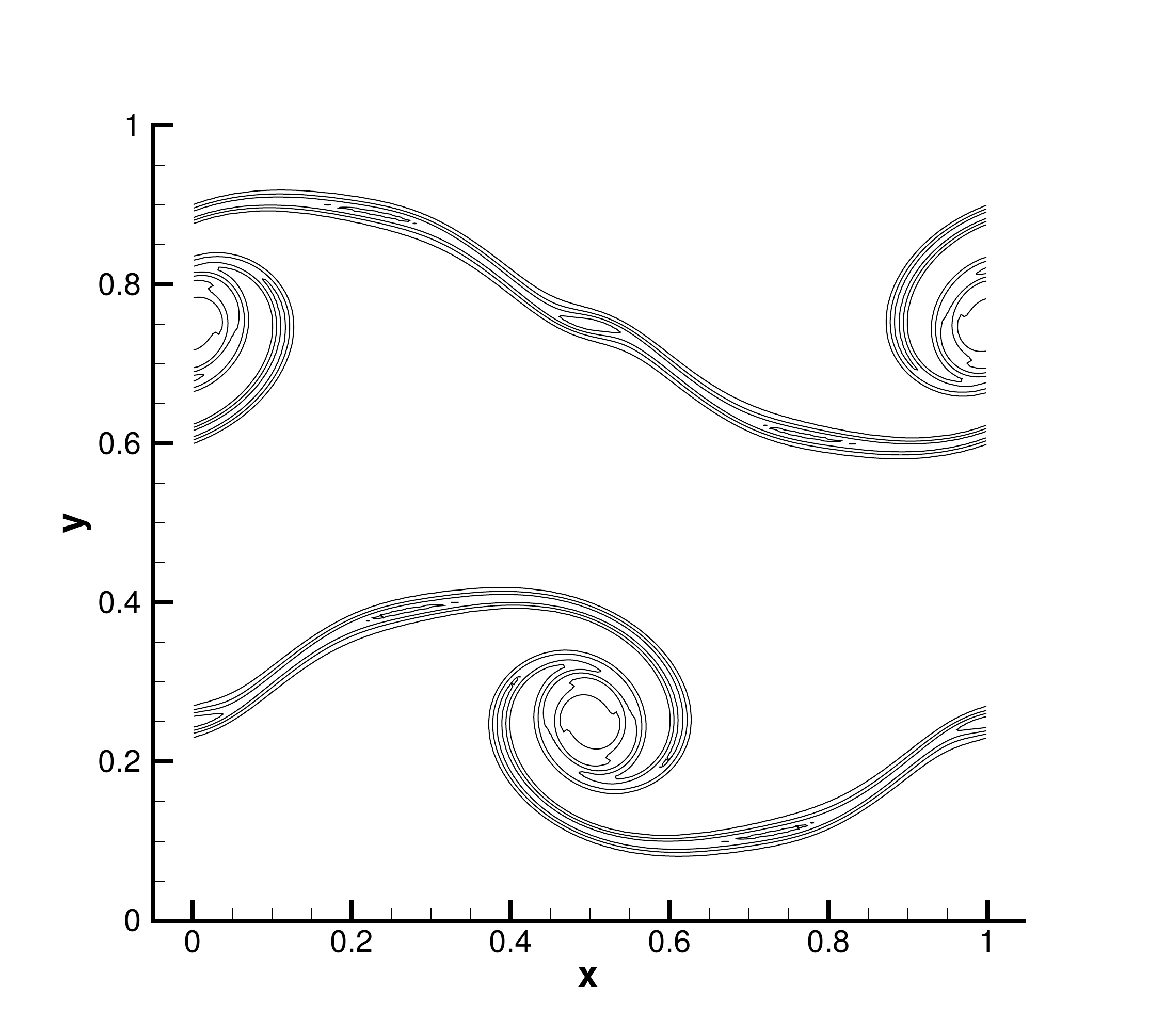}
	}				
	\caption{Two-dimensional double shear flow by WENO5-AO-GKS scheme (left) and WENO5-AO-HLLC scheme (right) : vorticity. The CFL number is 0.5, the output time is $t=0.8$, and the mesh number is $256\times256$. In figures, there are 10 equally spaced contours from -60 to 60.}
	\label{Double shear flow by GKS and HLLC}
\end{figure}

The vorticity contours $ \Omega = ( \frac{\partial V}{\partial x}-\frac{\partial U}{\partial y} )$ at $t=0.8$ obtained by WENO5-AO-GKS and WENO5-AO-HLLC are presented in Figure \ref{Double shear flow by GKS and HLLC}. The results show that the vortex in the whole domain is captured by WENO5-AO-GKS. From the results of WENO5-AO-HLLC, the prominent vortex structures are well resolved and the spurious roll-ups appear, especially in the region near the location $\left(0.5,0.75\right)$. These results indicate that WENO5-AO-GKS scheme has a slightly higher resolution than WENO5-AO-HLLC scheme even with the same reconstruction.

\subsubsection{Viscous shock tube}
\begin{figure}[htbp]
	\centering		
	\subfigure{
		\includegraphics[height=2.55cm]{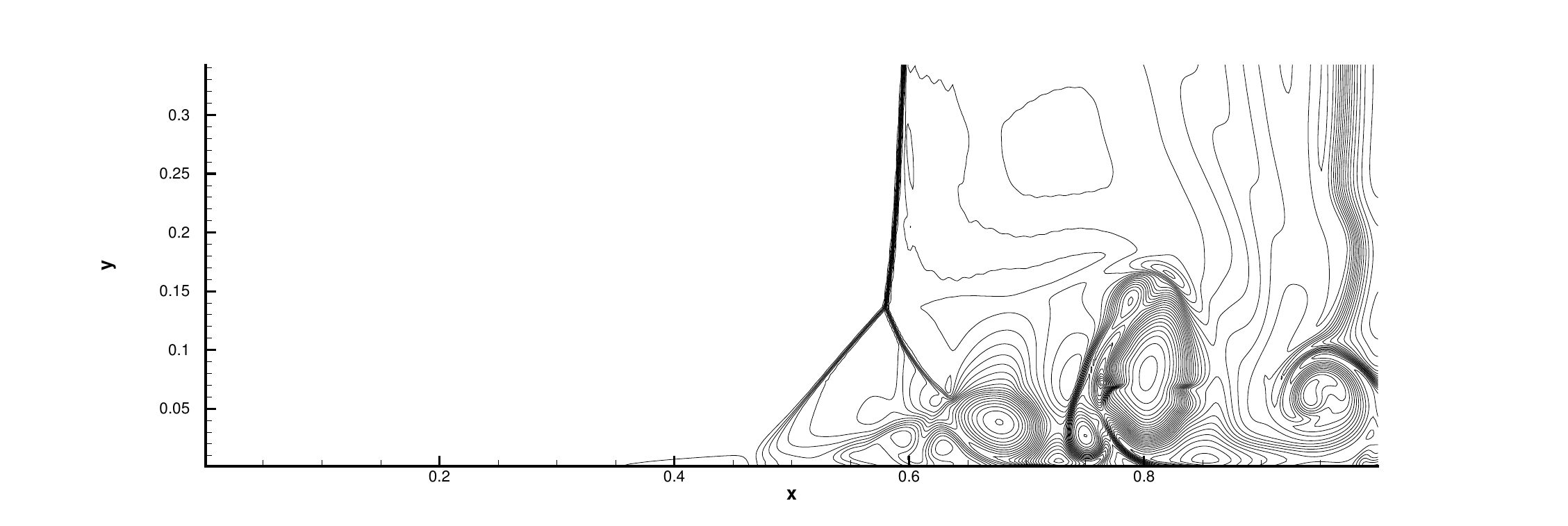}
	}
	\quad		
	\subfigure{
		\includegraphics[height=2.55cm]{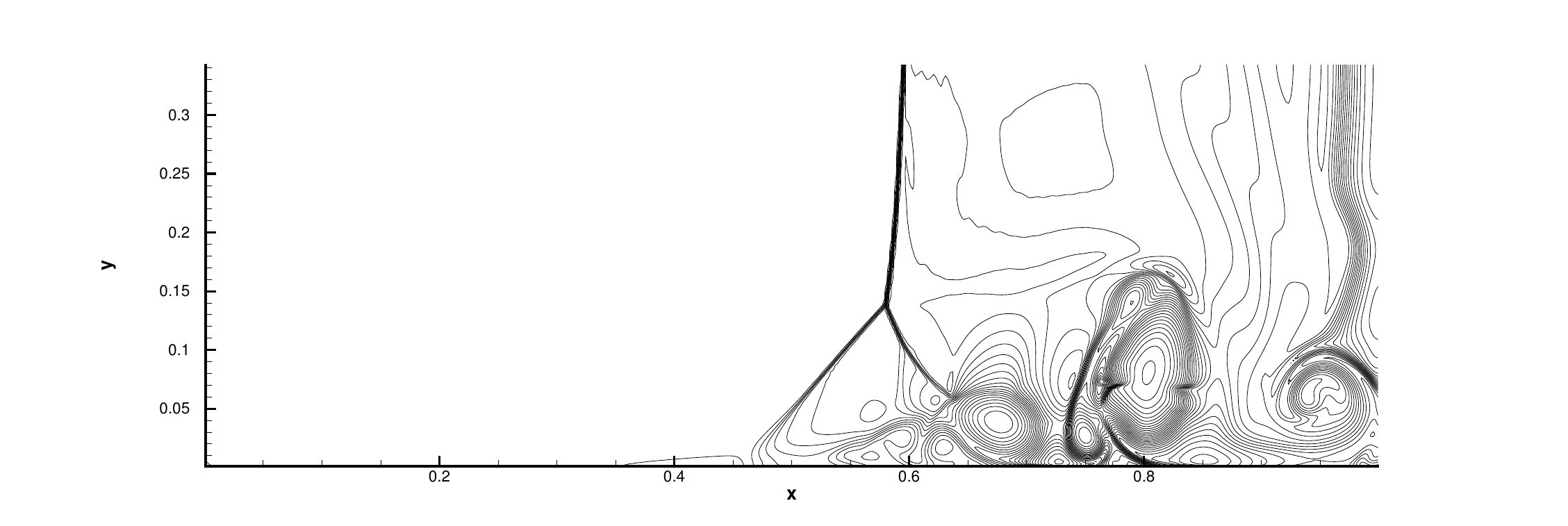}	
	}
	
	\subfigure{
		\includegraphics[height=2.55cm]{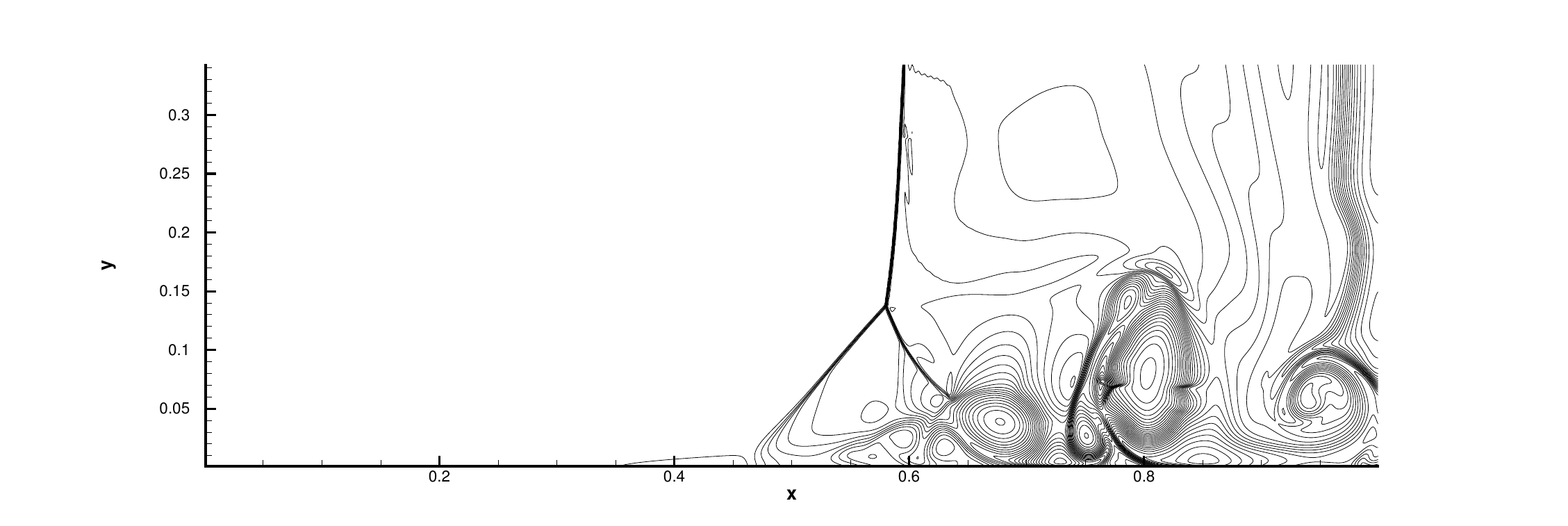}
	}
	\quad		
	\subfigure{
		\includegraphics[height=2.55cm]{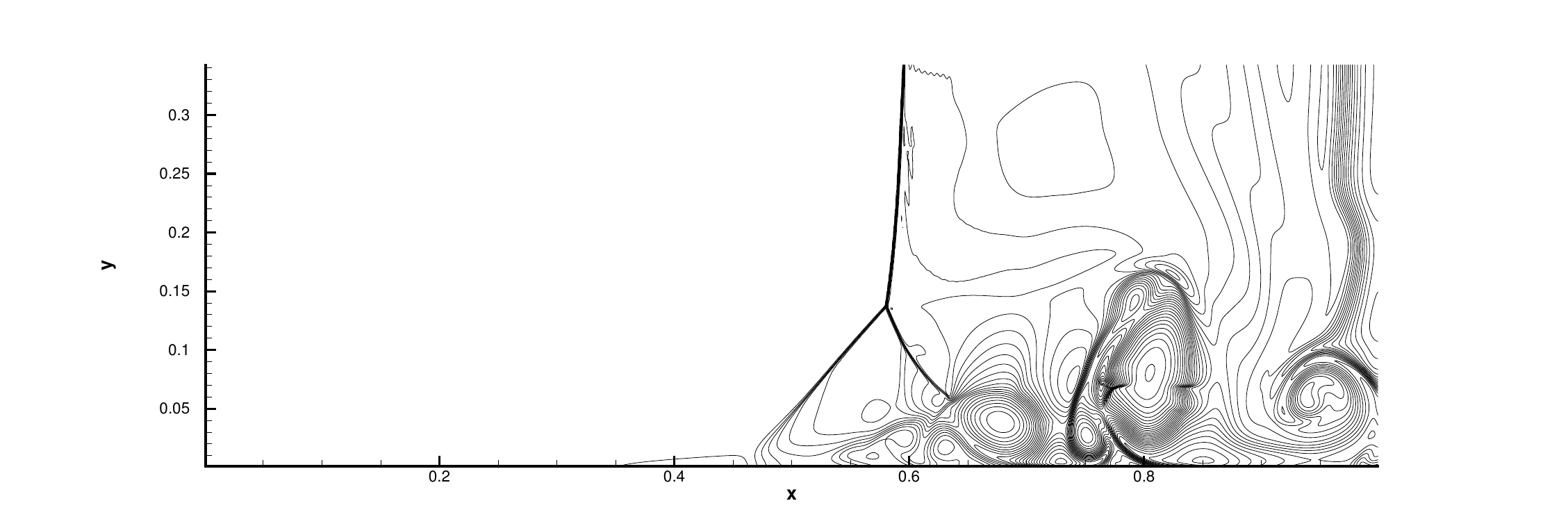}	
	}	
	\caption{Viscous shock tube problem with $Re=200$ by WENO5-AO-GKS scheme (left) and WENO5-AO-HLLC scheme (right): density distribution. For all cases, the CFL number is 0.2. For the top two figures, the mesh number is $500\times250$; and for the bottom two figures, the mesh number is $1000\times500$. For all figures, there are 30 equally spaced contours from 20 to 130.}
	\label{Viscous shock tube Re 200 by GKS and HLLC}
\end{figure}

\begin{figure}[htbp]
	\centering
	\subfigure{
		\includegraphics[height=6.5cm]{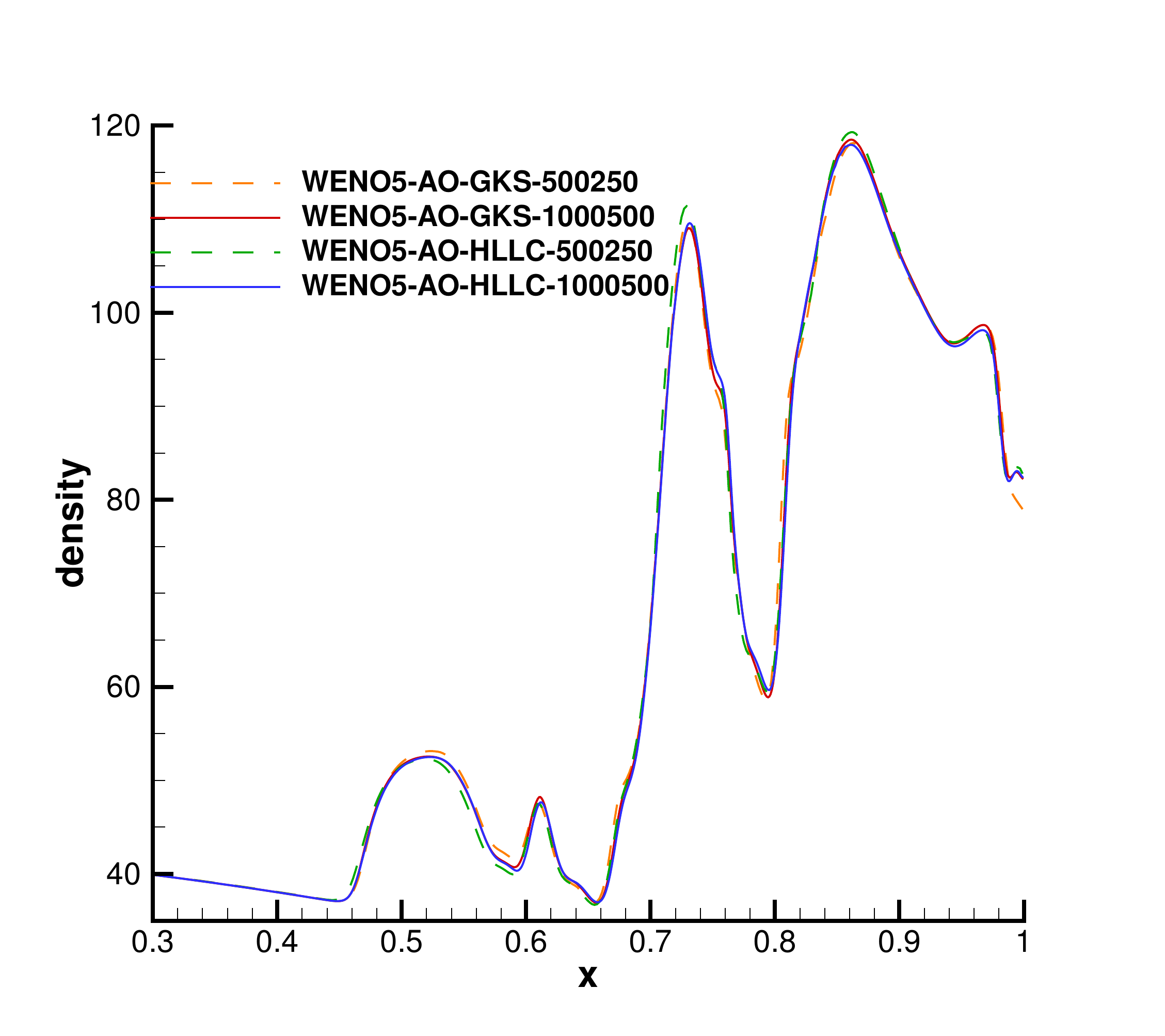}	
	}
	\quad
	\subfigure{
		\includegraphics[height=6.5cm]{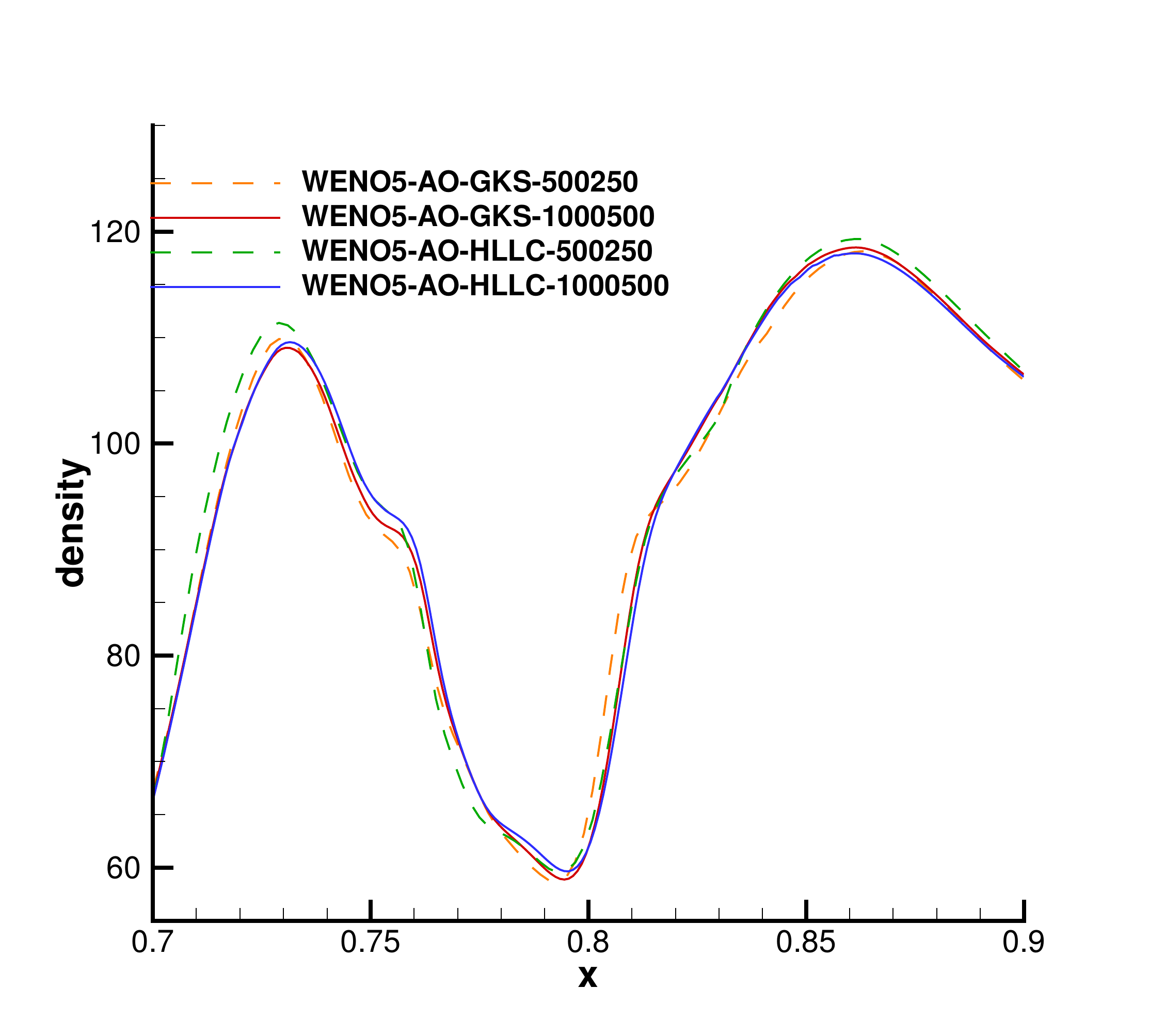}	
	}		
	\caption{Viscous shock tube problem of $Re=200$ by WENO5-AO-GKS scheme and WENO5-AO-HLLC scheme: density profile along the bottom wall ($y=0$). The right is the enlarged figure. For all cases, the CFL number is 0.2.}
	\label{Viscous shock tube Re 200 density profile by GKS and HLLC}
\end{figure}

Viscous shock tube problem is a viscous flow problem with a strong shock \cite{Viscousshocktube2000daru}. The interaction of reflected shock from the right wall and the viscous boundary layer produces a series of complex flow structures, such as the typical $\lambda-$shape shock configuration.
The initial condition is given by,
\begin{equation*}
(\rho,U,V,p)=\left\{\begin{aligned}
&(120, 0, 0, 120/\gamma),  & 0 < x \leq 0.5,\\
&(1.2, 0, 0, 1.2/\gamma),  & 0.5 < x < 1,
\end{aligned} \right.
\end{equation*}
where $\gamma=1.4$, $Pr = 0.73$.
The computational domain is $[0,1]\times[0,0.5]$. The simulation at $Re=200$ is tested.
The output time is $t=1.0$. For the boundary condition, the upper boundary is asymmetric boundary, and others are the non-slip adiabatic wall.
The density contours at $Re=200$ are shown in Figure \ref{Viscous shock tube Re 200 by GKS and HLLC}, where both WENO5-AO-GKS and WENO5-AO-HLLC can capture the main flow structures. The $\lambda-$shape structure, the vortices within the boundary layer, and the slip line in the lower right region are captured clearly.  The density profiles along the bottom wall are shown in Figure \ref{Viscous shock tube Re 200 density profile by GKS and HLLC} with the local enlargement.
The results show that both schemes have similar resolution. The results of WENO5-AO-GKS on both coarse and fine meshes seem to be closer than those of WENO5-AO-HLLC.

\subsection{3-D tests}

\subsubsection{Accuracy test}

The three-dimensional advection of density perturbation is adopted for accuracy test. The initial condition is,
\begin{gather*}
\rho(x,y,z)=1+0.2\text{sin}(\pi(x+y+z)),\\
U(x,y,z)=1, V(x,y,z)=1, W(x,y,z)=1, p(x,y,z)=1.
\end{gather*}				
The computational domain covers $[0,2]\times[0,2]\times[0,2]$. Under the periodic boundary condition, the analytic solution is as follows,
\begin{gather*}
\rho(x,y,z,t)=1+0.2\text{sin}(\pi(x + y + z - t)), \\
U(x,y,z,t)=1, V(x,y,z,t)=1, W(x,y,z,t)=1, p(x,y,z,t)=1.
\end{gather*}
The $L^1$ error and convergence order of WENO5-AO-GKS and WENO5-AO-HLLC at $t=2.0$ are shown in Table \ref{3D accuracy test by WENO-GKS} and Table \ref{3D accuracy test by WENO-HLLC}, respectively. The results show that the convergence orders of both schemes are higher than $4$ in this test.
The WENO5-AO-GKS scheme shows a slightly less absolute error than WENO5-AO-HLLC at different CFL number and mesh size.
\begin{table}[!htbp]\scriptsize
	\centering	
	\resizebox{\textwidth}{!}
	{
		\begin{tabular}{c|cc|cc|cc}
			\hline
			CFL & \multicolumn{2}{c|}{0.20} &\multicolumn{2}{c|}{0.60} &\multicolumn{2}{c}{1.00}\\
			\hline
			Mesh					&$L^1$ Error &Order &$L^1$ Error	&Order  &$L^1$ Error    &Order  \\
			5$\times$5$\times$5		&4.574909e-02&		&	3.704474e-02&		&	5.481745e-02&		\\
			10$\times$10$\times$10	&2.234252e-03&	4.36&	1.665667e-03&   4.48&	3.741869e-03&	3.87\\
			20$\times$20$\times$20	&7.589204e-05&	4.88&	6.007786e-05&	4.79&	2.052167e-04&	4.19\\
			40$\times$40$\times$40	&2.470600e-06&	4.94&	2.640903e-06&	4.51&	1.220770e-05&	4.07\\
			80$\times$80$\times$80	&8.596794e-08&	4.84&	1.424737e-07&	4.21&	7.525234e-07&	4.02\\
			\hline				
		\end{tabular}
	}
	\caption{3-D accuracy test: $L^1$ Error and convergence order by WENO5-AO-GKS scheme with different CFL numbers.}
	\label{3D accuracy test by WENO-GKS}		
\end{table}

\begin{table}[!htbp]\scriptsize
	\centering	
	\resizebox{\textwidth}{!}
	{
		\begin{tabular}{c|cc|cc|cc}
			\hline
			CFL & \multicolumn{2}{c|}{0.20} &\multicolumn{2}{c|}{0.60} &\multicolumn{2}{c}{1.00}\\
			\hline
			Mesh					&$L^1$ Error &Order &$L^1$ Error	&Order  &$L^1$ Error    &Order  \\
			5$\times$5$\times$5		&5.960514e-02&		&	6.192342e-02&		&	7.703808e-02&		\\
			10$\times$10$\times$10	&3.390217e-03&	4.14&	3.658550e-03&	4.08&	5.859740e-03&	3.72\\
			20$\times$20$\times$20	&1.209088e-04&	4.81&	1.312929e-04&	4.80&	2.588044e-04&	4.50\\
			40$\times$40$\times$40	&3.900971e-06&	4.95&	4.563747e-06&	4.85&	1.318523e-05&	4.29\\
			80$\times$80$\times$80	&1.283506e-07&	4.93&	1.845443e-07&	4.63&	7.688684e-07&	4.10\\
			\hline				
		\end{tabular}
	}
	\caption{3-D accuracy test: $L^1$ Error and convergence order by WENO5-AO-HLLC scheme with different CFL numbers.}
	\label{3D accuracy test by WENO-HLLC}		
\end{table}

\subsubsection{Compressible isotropic turbulence}
The decaying compressible isotropic turbulence is a case to evaluate the robustness of different schemes \cite{DecayingCIT2001Pullin, GKS-turbulence-CIT-Cao2019}. The definitions of flow variables are introduced first. The turbulent fluctuating velocity $U^\prime$ is,
\begin{equation*}
U^\prime=\left\langle \frac{U_1^2 + U_2^2 + U_3^2}{3} \right\rangle^{1/2},
\end{equation*}
where $\left\langle \cdots \right\rangle$ means the space average over the whole computation domain. Then, turbulence Mach number $Ma_t$ is given  by,
\begin{equation*}
Ma_t=\frac{\left\langle U_1^2 + U_2^2 + U_3^2 \right\rangle^{1/2}}{C} = \frac{\sqrt{3}U^\prime}{C},
\end{equation*}
where $C$ is the local sound speed.
Taylor microscale $\lambda$ is defined by,
\begin{equation*}
\lambda^2=\frac{\left(U^\prime\right)^2}{\left\langle \left( \partial U_1/\partial x_1 \right)^2 \right\rangle},
\end{equation*}
and the corresponding Taylor Reynolds number $Re_\lambda$ is
\begin{equation*}
Re_\lambda = \frac{\rho U^\prime \lambda}{\mu},
\end{equation*}
where $\mu$ is the dynamic viscosity coefficient determined by
\begin{equation*}
\mu = \mu_0 \left( \frac{T}{T_0} \right) ^{0.76}.
\end{equation*}
In this case, the velocity spectrum is given by,
\begin{equation*}
E(k) = A_0 k^4 e^{( -2k^2/k_0^2)} ,
\end{equation*}
where $A_0$ is the initial kinetic energy, $k$ is the wave number, and $k_0$ is the peak value of $k$. The initial turbulent kinetic energy $K_0$ and the initial large-eddy-turnover time $\tau_0$ can be obtained as follows,
\begin{gather*}
K_0 = \frac{3A_0}{64} \sqrt{2\pi} k_0^5, \\
\tau_0 = \sqrt{\frac{32}{A_0}} \left( 2\pi \right)^{1/4} k_0^{-7/2} .
\end{gather*}
The kinetic energy $K(t)$ and root-mean-square of density fluctuation $\rho_{rms} (t)$ are defined as
\begin{gather*}
K(t) = \frac{\left\langle \rho U_1^2 + \rho U_2^2 + \rho U_3^2 \right\rangle}{2},\\
\rho_{rms} (t) = \sqrt{ \left\langle \left( \rho - \overline{\rho} \right)^2 \right\rangle } .
\end{gather*}
In this case, there are strong shocklets and shock-vortex interactions in the flow field, especially at a high turbulence Mach number $Ma_t$. Therefore, it is challenging for high-order scheme to simulate high $Ma_t$ flow.
The simulations will cover a wide range of $Ma_t$ to compare the robustness of WENO5-AO-GKS and WENO5-AO-HLLC.
The mesh adopted in this case is $128^3$. Other parameters take the values $Re_\lambda = 72$, $A_0 = 1.3 \times 10^{-4}$, and $k_0 = 8.0$.

\begin{figure}[htbp]
	\centering		
	\subfigure{
		\includegraphics[height=6.5cm]{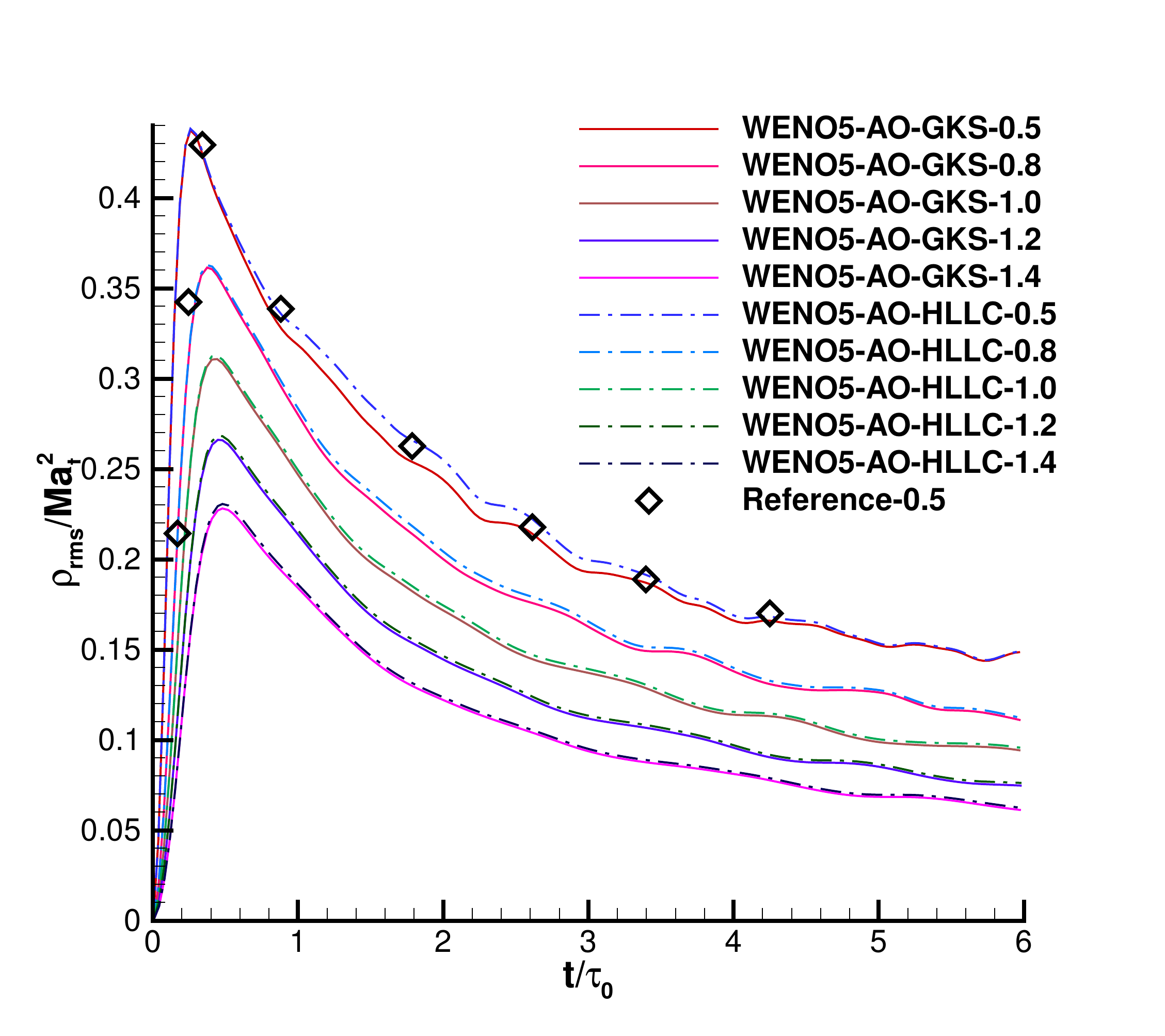}
	}
	\quad		
	\subfigure{
		\includegraphics[height=6.5cm]{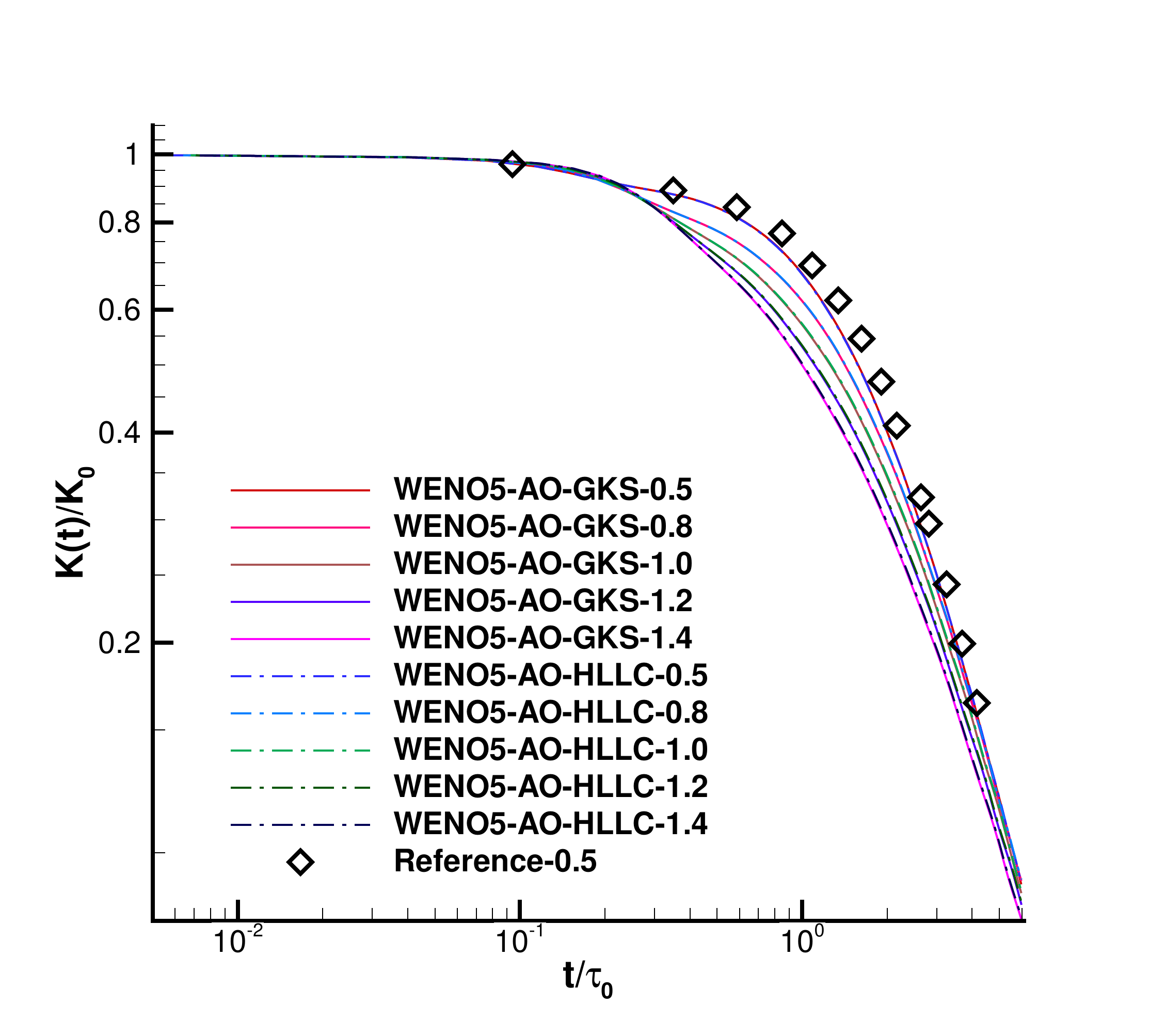}	
	}	
	\caption{Compressible isotropic turbulence at different $Ma_t$ by WENO5-AO-GKS scheme and WENO5-AO-HLLC scheme. The normalized kinetic energy (left) and normalized root-mean-square of density fluctuation (right). The CFL number is 0.3 for both WENO5-AO-GKS scheme and WENO5-AO-HLLC scheme. For all cases, the mesh number is $128^3$.}
	\label{CIT}
\end{figure}

\begin{figure}[htbp]
	\centering		
	\subfigure{
		\includegraphics[height=6.5cm]{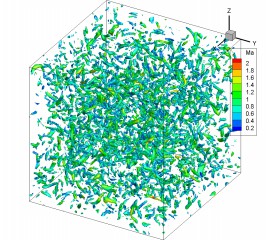}}
	\quad		
	\subfigure{
		\includegraphics[height=6.5cm]{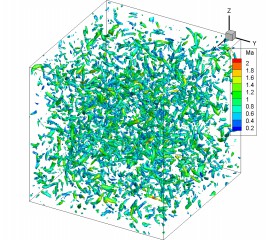}	
	}	
	\caption{Compressible isotropic turbulence with $Ma_t=1.4$: iso-surface of the second invariant of velocity gradient tensor $Q=25$ colored with local Mach number by WENO5-AO-GKS scheme (left) and WENO5-AO-HLLC scheme (right). The mesh number is $128^3$ and output time is $t/\tau_0=1.0$.}
	\label{CIT-Q25}
\end{figure}

The time history of normalized kinetic energy $K(t)/K_0$ and root-mean-square of density fluctuation $\rho_{rms}(t)/Ma_t^2$ are shown in Figure \ref{CIT}.
Both WENO5-AO-GKS and WENO5-AO-HLLC perform well for a wide range of $Ma_t$ from 0.5 to 1.4. The reference data of $Ma_t=0.5$ is obtained in \cite{DecayingCIT2001Pullin}.
When $Ma_t=1.4$, iso-surface of the second invariant of velocity gradient tensor $Q=25$ colored by the local Mach number at $t/\tau_0=1.0$ is shown in Figure \ref{CIT-Q25}. The results obtained by two schemes are nearly the same.
The CFL number is $0.3$ for both WENO5-AO-GKS and WENO5-AO-HLLC. But, WENO5-AO-GKS can take a larger CFL number 0.5 while 0.3 is the limit for WENO5-AO-HLLC in this case.
When the conservative flow variables at the interface for viscous fluxes are obtained by sixth-order central difference method, the WENO5-AO-HLLC can only work for $Ma_t$ up to 0.6, which is much smaller than $1.4$.
In the simulations, to improve the robustness of WENO5-AO-HLLC scheme, the conservative variables at the cell interface $Q_{i+1/2}$ are obtained by simple averaging of the left and right interface values of WENO5-AO reconstruction.
The above results show that WENO5-AO-GKS is more robust than WENO5-AO-HLLC in this case.

\subsubsection{Taylor-Green vortex}
The three-dimensional Taylor-Green vortex is studied by WENO5-AO-GKS and WENO5-AO-HLLC. The computational domain is $\left[-\pi L,\pi L\right] \times \left[-\pi L,\pi L\right] \times \left[-\pi L,\pi L\right]$, and the initial condition is
\begin{gather*}
U = U_0 \text{sin}\left(x/L\right)\text{cos}\left(y/L\right)\text{cos}\left(z/L\right),\\
V =-U_0 \text{cos}\left(x/L\right)\text{sin}\left(y/L\right)\text{cos}\left(z/L\right),\\
W = 0,\\
p = p_0 + \rho_0 U_0^2 \left(\text{cos}\left(2x/L\right)+\text{cos}\left(2y/L\right)\right)\left(\text{cos}\left(2z/L\right)+2\right) /16.
\end{gather*}
The simulation has $L=1$, $U_0=1$, $\rho_0=1$, and the Reynolds number $Re = U_0 L /\nu = 280$.
 The Mach number is $Ma=U_0/C$=0.1 and the sound speed is $C=\sqrt{\gamma R T}$. The mesh number is $64^3$, and periodic boundary condition is imposed at all boundaries.
The volume-averaged kinetic energy is defined as,
\begin{equation*}
E_k = \frac{1}{\rho_0 \Omega} \int_{\Omega} \frac{\rho \left(U^2+V^2+W^2\right)}{2}  \text{d}\Omega,
\end{equation*}
where $\Omega$ is the total volume of flow field. Besides, the dissipation rate of the kinetic energy is given by
\begin{equation*}
\epsilon_k = - \frac{\text{d}E_k}{\text{d}t}.
\end{equation*}

\begin{figure}[htbp]
	\centering		
	\subfigure{
		\includegraphics[height=6.5cm]{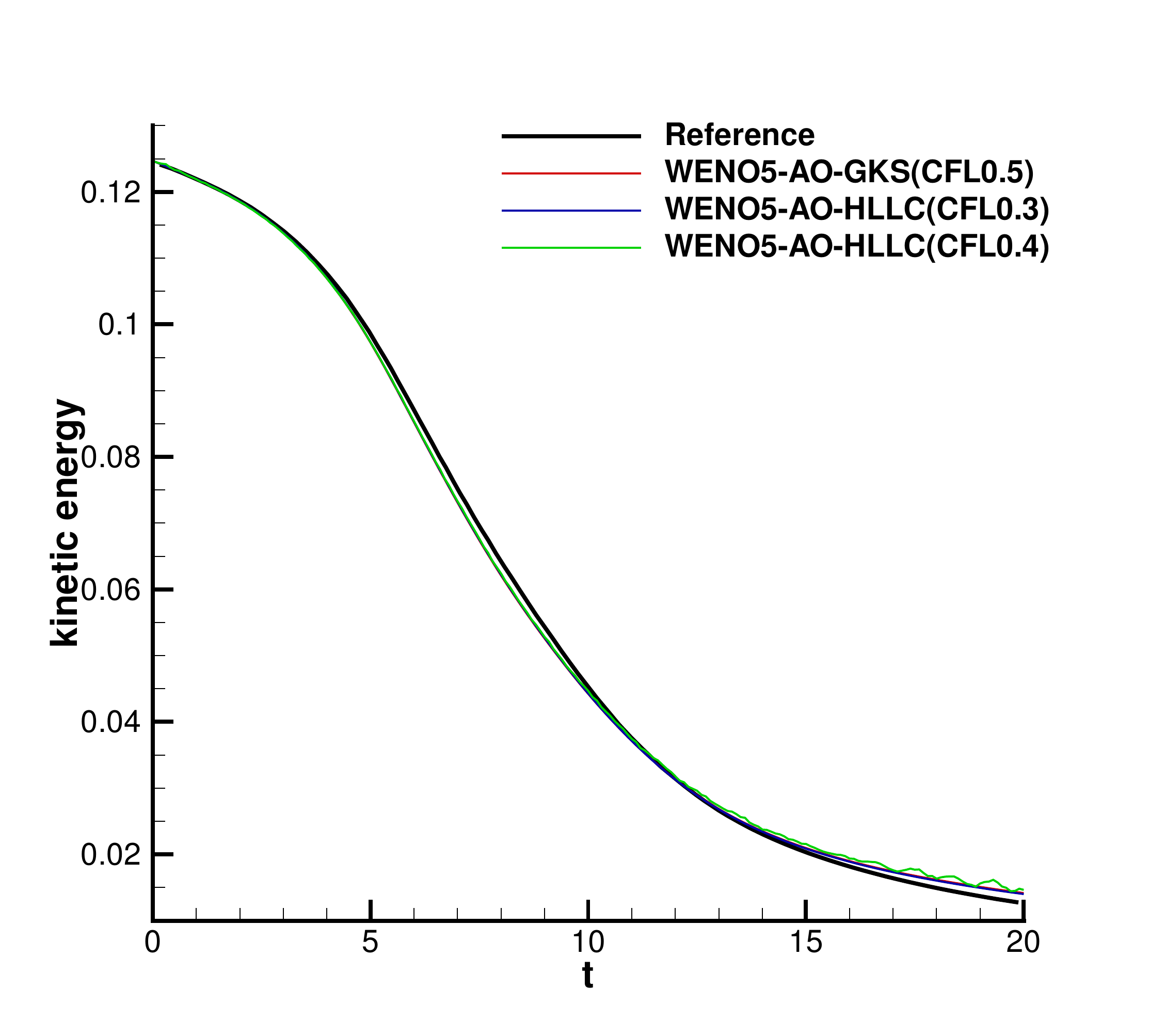}
	}
	\quad		
	\subfigure{
		\includegraphics[height=6.5cm]{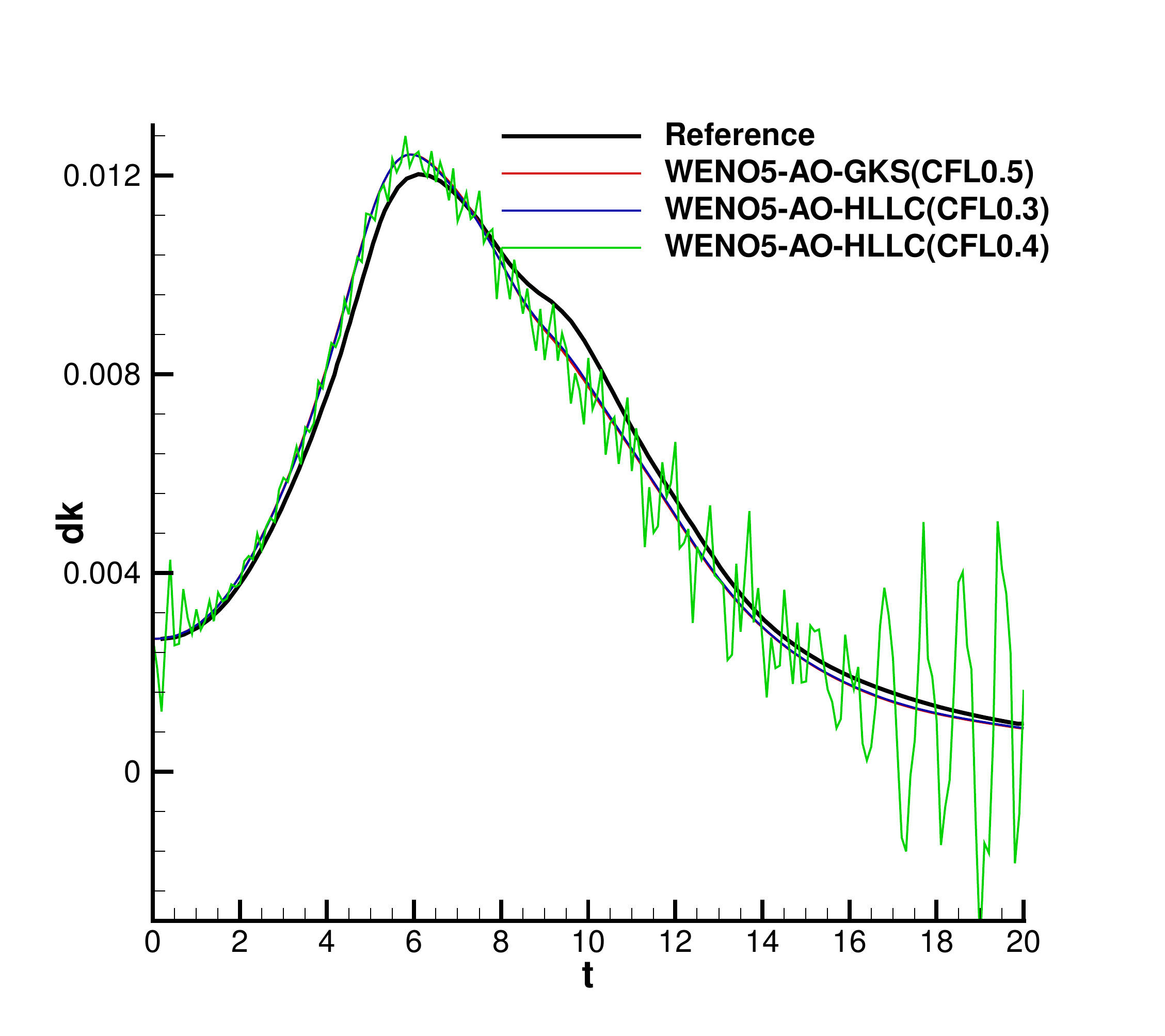}	
	}	
	\caption{Taylor-Green vortex problem with $Re=280$ by WENO5-AO-GKS scheme and WENO5-AO-HLLC scheme: the kinetic energy (left) and the dissipation rate (right). The CFL number is 0.5 for WENO5-AO-GKS scheme and 0.3 for WENO5-AO-HLLC scheme. For both cases, the mesh number is $64^3$.}
	\label{Taylor-Green vortex}
\end{figure}

The linear reconstruction is taken for both schemes in this test case. The results are presented in Figure \ref{Taylor-Green vortex}, and are compared with the reference solution of \cite{S2O4-threedimensional-Pan2018}. The CFL number is $0.5$ for WENO5-AO-GKS while $0.3$ for WENO5-AO-HLLC. When CFL number is $0.4$, WENO-AO-HLLC will generate large oscillation.

\section{Computational efficiency}
The computational efficiency of WENO5-AO-GKS and WENO5-AO-HLLC is compared in 2-D and 3-D cases.
For both schemes, the main computational cost includes two parts, reconstruction, and evolution.
For the reconstruction, WENO5-AO-HLLC needs only pointwise conservative variables, while the derivatives are also needed in WENO-AO-GKS.
However, additional reconstruction through central difference method for the viscous terms is required in WENO5-AO-HLLC.
For the evolution stage, the GKS flux is more expensive than HLLC, but GKS uses two stages instead of four stages in HLLC to achieve 4th-order time accuracy.

The viscous shock tube is used to test the computational efficiency. The mesh points in the test are 1000$\times$500.
The viscous flux in WENO-AO-HLLC is obtained through sixth order central difference method, where the inviscid and viscous terms are coupled in the
GKS flux.
The WENO5-AO reconstruction is based on characteristic variables for both schemes.
In this case, the computation time and the relative efficiency are listed in Table \ref{2D computational efficiency test for viscous shock tube}. The computation times shown in Table \ref{2D computational efficiency test for viscous shock tube} are obtained for $10$ time steps by a single Intel core i7-9700 @ 3.00GHz. The results show that WENO5-AO-HLLC  is 27\% more expensive than WENO5-AO-GKS in the 2-D viscous problem.
The next test is the compressible isotropic turbulence in 3-D.
Again, the WENO5-AO reconstruction is based on characteristic variables for both schemes.
The computational time is collected by running the code for 10 time steps, and the results are shown in Table \ref{3D computational efficiency test for CIT}. The calculation time of WENO5-AO-HLLC is about 15\% less than WENO5-AO-GKS. This is mainly due to the three-dimensional reconstruction, where the reconstruction in two tangential directions on both sides of a cell interface is needed in WENO-AO-GKS, instead of one tangential direction in 2D case. In this test, WENO-AO-GKS can take a CFL number 0.5, and WENO5-AO-HLLC can take a CFL number 0.3 only.
As a result, WENO5-AO-GKS can have a slightly better overall efficiency in 3D case.

\begin{table}[!htbp]\small
	\centering		
	{
		\begin{tabular}{c|c|c}
			\hline
					      & CPU time ($s$) & Time ratio\\
			\hline			
			WENO5-AO-GKS  & 154.91 & 1.00\\
			WENO5-AO-HLLC & 196.47 & 1.27\\		
			\hline
		\end{tabular}
	}
	\vspace{-2mm}\caption{2-D computational efficiency test of viscous shock tube problem. The mesh number is 1000$\times$500. The shown CPU time is obtained for 10 time steps by a single Intel core i7-9700 @ 3.00GHz. }
	\label{2D computational efficiency test for viscous shock tube}		
\end{table}

\begin{table}[!htbp]\small
	\centering
	{
		\begin{tabular}{c|c|c}
			\hline
					      & CPU time ($s$) & Time ratio\\
			\hline			
			WENO5-AO-GKS  & 476.04 & 1.00\\
			WENO5-AO-HLLC & 403.03 & 0.85\\		
			\hline
		\end{tabular}
	}
	\vspace{-2mm}\caption{3-D computational efficiency test of compressible isotropic turbulence problem with $Ma_t=0.5$ and CFL = 0.3. The mesh number is $128^3$.The shown CPU time is obtained for 10 time steps by a single Intel core i7-9700 @ 3.00GHz.}
	\label{3D computational efficiency test for CIT}		
\end{table}

\section{Conclusion}

A comparison of performance for two high-order schemes, namely WENO5-AO-GKS and WENO5-AO-HLLC, is presented.
Both schemes use the fifth-order WENO-AO reconstruction, the differences are mainly coming from the flux functions and the temporal updating schemes.
In GKS, due to the time accurate flux and its time derivative the multistage and multiderivative (MSMD) is used to update the solution.
The two-stage fourth-order temporal discretization achieves a 4th-order temporal accuracy. For HLLC, four stages Runge-Kutta method
is used for the time accuracy.
In WENO-AO-GKS, both inviscid and viscous flux terms can be evaluated from a single time-dependent gas distribution function.
In WENO5-AO-HLLC, HLLC provides inviscid flux and a sixth-order central difference method is used to discretize the viscous flux.
In the 3D accuracy test, both schemes can achieve the expected order of accuracy, and WENO5-AO-GKS shows a slightly smaller absolute $L^1$ error.
In terms of the shock and contact wave capturing, both schemes perform well and have similar robustness.
With the same mesh and CFL number, WENO5-AO-GKS shows better accuracy in the double shear layer test.
In the Noh problem, WENO5-AO-GKS presents favorable robustness.
For the compressible isotropic turbulence and three-dimensional Taylor-Green vortex problem,
WENO-AO-GKS can take a large time step with CFL number 0.5, instead of 0.3 for WENO5-AO-HLLC.
For two-dimensional viscous shock tube problems, WENO5-AO-HLLC is (27\%) more expensive than WENO-AO-GKS.
While for the three-dimensional viscous test, WENO5-AO-HLLC is (15\%) more efficient than WENO5-AO-GKS.
WENO-AO-GKS requires the reconstruction of flow variables in the normal and two tangential directions on both sides of a cell interface in the 3D case.
The multi-dimensional property and the coupling of inviscid and viscous fluxes in WENO5-AO-GKS have obvious 
advantages when the scheme is extended to the flow computation with unstructured mesh.

\bibliographystyle{plain}%
\bibliography{jixingbib1}
\end{document}